# Coma Morphology, Numerical Modeling, and Production Rates for Comet C/Lulin (2007 N3)

Allison N. Bair[1], David G. Schleicher[1], and Matthew M. Knight[2]
Contacting author: bair@lowell.edu

ABSTRACT

We report on narrowband photometry and extensive imaging observations of comet C/Lulin (2007 N3) obtained at Lowell Observatory during 2008 and 2009. Enhanced CN images revealed a double corkscrew morphology with two near-polar jets oriented approximately east-west, and both CN and dust images showed nightly rotational variability and seasonal changes in bulk morphology. We determined a rotational pole direction of RA/Dec = 81°/+29° with an obliquity of 97°, and a sidereal rotation period of 41.45 ± 0.05 hr. Monte Carlo numerical modeling best replicated the observed CN features with an eastern source area at lat/long -80°/125° and a ~10° radius and a western source area at lat/long +77°/245° and a ~20° radius, ~4× larger than the eastern source. An additional small, near-equatorial source was necessary to reproduce some dust features. Water morphology, based on OH, was quite different than that of the carbon-bearing species, implying a different driver for the polar jets such as CO or $CO_2$. Ion tails were detected in decontaminated images from both the dust and NH filters, likely being $H_2O^+$ and $OH^+$, respectively. We measured water production both before and after perihelion, and extrapolate peak water production at perihelion to be about $1.0 \times 10^{29}$ molecules s$^{-1}$. We estimate an active fraction of only 4-5% and a nucleus radius of up to ~8 km. Our data suggest that Lulin, defined as dynamically new in a statistical sense, behaves more like a long-period comet due to its nearly asteroidal early appearance, isolated source regions, and dust properties.

## 1. INTRODUCTION

Comet C/Lulin (2007 N3) was first identified on images obtained for the Lulin Sky Survey on 2007 July 11, when the comet was at a heliocentric distance of 6.4 AU (Young 2007). Lulin reached perihelion on 2009 January 11 – exactly 18 months after its discovery – and briefly became a naked-eye object six weeks later as it passed just 0.41 AU from the Earth. Orbital integrations indicated that its original reciprocal semi-major axis $1/a_0$ value of $31 \times 10^{-6}$ AU$^{-1}$, prior to planetary perturbations, placed it into the dynamically new class of comets. This implied Lulin was on its first passage into the planetary regime, therefore having an unevolved nucleus. With an orbital inclination of 178.4°, Lulin was traveling nearly in Earth's orbital plane but in the opposite direction, resulting in an exceedingly short interval when the comet was reasonably bright and within about 1 AU of Earth. During this brief time, there was a massive change in viewing geometry culminating in Lulin reaching a phase angle of 0.1° on February 26. This favorable viewing geometry, combined with its bright apparition, made Lulin an excellent candidate for study at a variety of wavelengths and with a variety of techniques.

Several papers have resulted from these efforts, and many more preliminary results were presented at meetings or in IAU Circulars. Production rates and/or compositional studies, mostly by means of spectroscopy, were obtained from the UV and visible (Combi et al. 2009a,b; Bodewits et al. 2011; Carter et al. 2012; Morgenthaler et al. 2009), to the near-IR (Woodward et al. 2011, Gibb et al. 2012, Kobayashi et al. 2009; Ootsubo et al. 2010), and the mm and radio (Biver et al. 2009; Charnley et al. 2009; Lovell & Howell 2009). Photometry of the dust was obtained (Joshi et al. 2011) along with polarimetry in both the visible and IR (Woodward et al. 2011), while imaging in the visible was also acquired (Hicks et al. 2009). Our own initial discoveries from narrowband imaging, including Lulin's pair of jets, its long rotational period, and a preliminary pole solution were announced shortly after perigee, while additional results from photometry, imaging and modeling were reported at DPS meetings (Knight & Schleicher 2009a,b; Bair & Schleicher 2016).

In this paper we present detailed results of our extensive observing campaign on Comet Lulin. We use narrowband optical imaging obtained on 32 nights during the first half of 2009, in which we identified two side-on CN jets whose corkscrew morphology repeated with an apparent rotational period of 42.0±0.5 hr. Numerical modeling, combined with the wide range of viewing geometry at which Lulin was observed, allowed us to develop a 3-D model of the orientation of the nucleus and the location and extent of its two major active regions, yielding gas outflow velocities and a precise sidereal rotation period. In addition to the imaging, we have a limited amount of narrowband photometry obtained over a longer time interval, providing Lulin's chemical composition and its production rates as a function of distance. Our observations and reductions are summarized in Section 2, while in Section 3 we discuss the observed physical characteristics of Lulin. We present our model solution in Section 4, our photometry results in Section 5, and discuss and summarize our results and their implications for other comets in Section 6.

## 2. OBSERVATIONS, REDUCTIONS, AND METHODOLOGIES

*2.1 Overview, Instrumentation, and Observations*

Due to our having multiple goals associated with Comet Lulin's unusual apparition, we utilized differing instrumentation during our observational campaign. All data were obtained with the John S. Hall 42 inch (1.1 m) telescope at Lowell Observatory. As noted in the Introduction, the extremely wide range of viewing geometry was due to Lulin's orbital inclination of 178°, thus being almost in the plane of the Earth's orbit but moving in the opposite direction. This also meant that the interval from one solar conjunction to the next was very rapid and, avoiding the bright moon as well, the useful duration of the primary apparition was only from late January to mid-May. The relatively close approach to Earth in late February of 2009 also meant that Lulin's peak brightness occurred then, rather than at perihelion (2009 Jan 10.64; heliocentric distance, $r_H$, = 1.212 AU); in fact, the comet was nearly 4 magnitudes brighter at perigee (Feb 24.3; geocentric distance, $\Delta$, = 0.411 AU) than at perihelion[1]. While the comet was too faint for narrowband imaging prior to the late 2008 solar conjunction, our other standard technique – narrowband photometry using a traditional photoelectric photometer – was feasible in summer 2008. Unfortunately, we only obtained one night of data in this time frame, as a bright star was within an arcminute of the comet on a second night, and the entire observing run the following month was weathered out. We also lost our scheduled photometer run in January, although subsequent monthly runs each yielded at least one night of photometry. In all, a total of 24 sets of photometry, each composed of OH, NH, CN, $C_3$, and $C_2$, along with three continuum points (ultraviolet (UV), blue, and green) from the HB comet filter sets (Farnham et al. 2000), were obtained over 6 nights; dates and associated values for other parameters are summarized in Table 1.

[TABLE 1 here; Photometry Observing Circumstances and Fluorescence Efficiencies for Comet Lulin (2007 N3)]

Imaging was obtained using a SITe 2048 × 2048 CCD camera and binned 2 × 2, resulting in an effective pixel scale of 1.18 arcsec per pixel and corresponding to about 350 km at perigee on Feb 24 but nearly 2400 km at the end of our observations in mid-May. Useful observations were obtained on a total of 32 nights, though ten of these nights only have "snap shots" taken by colleagues working on other projects. Along with a broadband Kron-Cousins *R* filter, various subsets of the HB filter set were used for the imaging each night, depending on Lulin's brightness, the number of hours it was available, and weather conditions. Having the best signal-to-noise (S/N) and contrast to the underlying continuum, the CN emission band is our primary gas filter, and we have found that it can frequently be used for morphological and rotational studies under non-photometric conditions (e.g., Knight & Schleicher 2011, Knight & Schleicher 2015). Specific filters employed each night are listed in Table 2, along with other values associated with the observing circumstances for the imaging. Note that, in an effort to obtain imaging on every night near the comet's closest approach to Earth in late February yet also obtain some conventional photometry, we swapped instruments during the night of Feb. 26. By coincidence the night of minimum phase angle, the value of the phase angle differed between the times for the two instruments, from 0.7° with the CCD to 0.07° with the photometer. Thus this night is listed in both Tables 1 and 2 but having different values.

---

[1] http://www.aerith.net/comet/catalog/2007N3/2007N3.html

[TABLE 2 here; CCD Observing Circumstances for Comet Lulin]

### *2.2 Photometer Reductions and Analyses*

We employed standard techniques, comet flux calibration stars (Farnham et al. 2000), and reduction coefficients (cf. A'Hearn et al. 1995; Schleicher & Bair 2011) to compute fluxes, column abundances within the photometer entrance aperture, $M(\rho)$, and production rates, $Q$, for each of the five daughter gas species. Water production rates, based on $Q$(OH), were also derived. Fluxes for each continuum filter – UV, blue, and green – were computed, along with the resulting proxy for dust production, $A(\theta)f\rho$ (A'Hearn et al. 1984). One-sigma uncertainties were computed based on the observational photon statistics. Due to the unusual geometry associated with Lulin remaining extremely close to the ecliptic, the phase angle not only changed significantly but dropped to 0.07° on Feb 26, two days following perigee. To correct for phase angle effects, we therefore also normalized all of the dust results to 0° phase angle using a composite dust phase curve (see Schleicher & Bair 2011). We defer presenting these narrowband photometry results to Section 5, along with other, associated results.

### *2.3 CCD Reductions and Analyses*

The CCD reductions followed standard techniques for bias removal and flat fielding. Absolute calibrations of narrowband filters on photometric nights were performed using the same HB standard stars as with the photometer (Farnham et al. 2000). *R*-band images were calibrated using the SDSS-R9 system (Ahn et al. 2012) transformed to *R* with on chip field stars using the Python package Photometry Pipeline (PP; Mommert 2017). Magnitudes for the *R*-band continuum were measured in a series of apertures having radii of $\log(\rho) = 4.0, 4.3, 4.6$, and $4.9$ km, after centroiding on the central condensation. The primary aperture used in our analyses reported here had the radius fixed in size at the comet of 40,000 km on all nights to best match the preferred photometer aperture (see Section 5) and minimize rotational variations when measuring $A(\theta)f\rho$; these broadband, on-chip calibrated frames were primarily used in measuring the phase angle effects presented in Section 5.4. When conditions permitted, $A(\theta)f\rho$ was also determined from the blue and red narrowband continuum filters. Although the majority of imaging nights were not photometric, fortunately the CN emission has sufficient contrast to the underlying continuum that we can easily enhance the flat fielded images and detect and measure CN morphological features, i.e. jets. The same is true for OH in principle, but the S/N is so much lower than that of CN that the comet's brightness (or lack thereof) is another major factor as to how many nights of useful OH imaging exists. However, $C_2$, $C_3$, and NH all require proper calibration and continuum subtraction to be useful, due to much lower contrast of the bands to the underlying continuum, and therefore we have far fewer "pure gas" images for these species.

We followed our normal methodology of removing the bulk radial fall-off by means of azimuthal median profile subtraction (cf. Knight & Schleicher 2015) to enhance images and reveal faint structures in the coma. This process is relatively benign, in that the location of peak flux is not altered and artifacts are not introduced. The comet was strongly centrally condensed, making it easy to centroid on the nominal nucleus position. As demonstrated next, upon applying the enhancement, two jets were revealed in nearly all of the gas images. Additional enhancements were also applied (Schleicher & Farnham 2004, Samarasinha & Larson 2014), but in general were less helpful for our investigations, and are not discussed further.

## 3. OBSERVED MORPHOLOGICAL PROPERTIES

We tracked Lulin post-perihelion during three main imaging runs surrounding opposition, and continued to monitor its activity until it became too faint three and a half months after we began. During this relatively short interval Lulin's elongation from the Sun, the best representation of the geocentric change in geometry, went from 79°, through 180°, and back down to 45°, for a total change of 236°. As noted in Section 2.3, our most intensive monitoring of the comet was performed with CN and broad-band *R* filters, while additional filters were used when time and circumstances allowed.

*3.1 Coma Morphology and Seasonal and Rotational Variability in CN and Dust*

Our raw images of Lulin showed an elongated coma. Once enhanced the CN images revealed two corkscrew jet features lined up approximately east to west. Each jet remained on its respective side of the coma from month to month, even with the massive change in projection effect, though we did observe the relative brightness of each jet to change over time. In the CN images, the west jet was dominant early on (Figure 1), but by our second run, beginning in late Feb, each jet appeared to be of similar strength, while alternating in brightness with the rotation of the nucleus (Figure 2). A similar appearance continued into our third run, though the overall brightness of the comet had dropped significantly by this time (Figure 3). It wasn't until our fourth observing run, beginning Apr 23 (102 days after perihelion; $\Delta T = +102$ day), that Lulin's east jet became dominant in the CN images.

[Figure 1 here; CN and Dust rotational sample images – Run #1]
[Figure 2 here; CN and Dust rotational sample images – Run #2]
[Figure 3 here; CN and Dust rotational sample images – Run #3]

Rotational variations were evident in the enhanced CN images from each of our three main observing runs, but were most pronounced during our second run when Lulin was both at its brightest and was available for most of the night. A full rotational cycle is given in Figure 2. The images show each corkscrew feature moved steadily outward during each night of observations, while the intensity of the CN jets alternated in brightness with the rotation of the nucleus. A similar behavior is seen in images from the preceding run (Figure 1), though the nightly observing window is much shorter and as a result our coverage is not as comprehensive. By the third run (Figure 3), it was difficult to distinguish outward movement within a single night, but we still observed the jets to alternate in brightness. No rotational variability could be distinguished within individual nights during the final run in late April. Overall, the basic situation was clear – the rotation axis was aligned ESE-WNW in projection throughout the 15-week interval, a source region was located near each pole of the nucleus, producing this pair of corkscrew features, and the bulk brightness variations through the apparition were likely due to a seasonal change in solar illumination.

Unlike the CN, the *R*-band, i.e. dust, features presented a much more complicated situation, with greater changes from month-to-month. While there are some similarities with CN in the first run, with extensions along the same axis to the ESE and WNW, the detailed structures are quite different. Moreover, by the second run, and continuing into the third and fourth runs, the only apparently common component is a bulk feature towards the ESE.

Ultimately, we concluded that there were several components that must be disentangled to understand the various structures in the *R*-band images, one of which is not even dust but rather an ion tail, based both on its morphological appearance and direction. Not only is it evident during both the first and second runs, but its orientation changed abruptly as we passed through minimum phase angle and the Sun switched sides (Figure 4). Since $H_2O^+$ is the most prevalent ionic feature within the *R*-band, we conclude that this is the most likely identification. Another feature evident in the first run is in the opposite direction from the ion tail, i.e. sunward, and it appears to be old tail material seen in projection far behind the comet, though modeling this feature is beyond our capabilities at this time. Throughout the second run a broader feature is present in the same location that we also, but more tentatively, identify as old tail material. By the third run a broad feature remains towards the east and is likely young tail material but perhaps overlapping older and more distant material. Carter et al. (2012) saw similar features in their dust images, and also concluded they consist of long-lived tail material.

[Figure 4 here; R-band and NH ion tail sets from Runs 1 and 2]

Much closer to the nucleus are additional dust features. In the first run, material appears to be emitted towards the S, SE, or E (depending on the rotational phase), and curl past the south and end up towards the west. However, we think this is actually the overlap of two features, the one to the west corresponding to material originating from the same source producing the western CN jet, but having a shorter spatial extent due to the slower velocity of dust grains, and the one emitted towards the S, SE, or E that curls past the south and then moves anti-sunward due to radiation pressure, eventually overlapping the western feature. By the second run, circumstances have changed again. There is no evidence of a western feature, but now there is a feature towards the SSE and another towards the NNW, with variations in relative intensities with rotational phase. By the third run, the enhanced images are dominated by material towards the east, and with the much greater concentration of stars (and poorer S/N), structures near the nucleus are difficult to discriminate.

Because modeling was required to disentangle the various near-nucleus dust features during the apparition, we defer additional discussion until Section 4.8.

### 3.2 Comparison of Morphology to the Other Gas Species

When observing circumstances allowed, we obtained images in additional narrowband filters to see how the other gas species behaved in comparison to CN and to dust (the dates and filters used are included in Table 2), but as already stated in Section 2.3, we needed photometric nights such that calibrations could be performed and underlying continuum could be removed. Our best images for comparison during the first observing run were taken on 2009 Feb 1, three weeks after perihelion ($\Delta T$ = +21 day; Figure 5 top). Our nightly observing window was relatively short, but we were able to obtain images for the pure carbon species, $C_2$ and $C_3$, in addition to our images monitoring CN and dust. As expected based on many previous comets that we've observed (i.e. Knight & Schleicher 2013), $C_3$ jets exhibit a distribution quite similar to those of CN but do not extend nearly as far from the nucleus due to $C_3$'s much shorter lifetime. In the best nights from our following runs, Feb 28 ($\Delta T$ = +48 day; Figure 5 middle) and Mar 29 ($\Delta T$ = +77 day; Figure 5 bottom), $C_3$ continues to behave as expected. In contrast, $C_2$ jets usually appear broader and more diffuse than CN jets, due to having multiple parents and at least one grandparent, thereby resulting in a wider range of excess energy with the various dissociations. While this is indeed the case during the second and third observing runs, the $C_2$ jets appear significantly narrower (i.e. more similar to CN) during the first run (Figure 5 top). A tentative explanation for the evolution in the appearance of $C_2$ is that during the first run the collision zone extended beyond where most $C_2$ parent dissociations took place, and that collisions effectively dampened out the usual vectorial addition of velocities caused by the excess energy of dissociation of the parent species.

[Figure 5 here; gas and dust images from run 1 (5a,), run 2 (5b), and run 3 (5c)]

OH emission presents a far different situation. Unfortunately, the short observing window and correspondingly high airmass during the first run precluded OH imaging. The second run, however, provided near-perfect circumstances, and the OH morphology appears nothing like the carbon-bearing species. The unenhanced OH images look very symmetric, but after enhancement faint features are seen. At most rotational phases, there is a broad feature generally towards the south, and only near phases 0.4-0.5 is the feature towards the east. A month later S/N was far worse due to the rapid brightness drop in Lulin, but it was apparent that the OH now had a spherically symmetric appearance, with no evident structure even with enhancement. While we will return to this behavior during modeling in Section 4, we simply note here that there have now been several comets, including Hartley 2 (Knight & Schleicher 2013), C/Jacques 2014 E2 (Knight & Schleicher 2014), 45P/Honda-Mrkos-Pajdusakova (unpublished), C/Catalina 2013 US10 (unpublished), and C/Lovejoy 2014 Q2 (unpublished), for which OH emission morphology is quite different than that of CN, $C_2$, and $C_3$, and Lulin is clearly another case of this.

Our biggest surprise occurred when we examined the continuum-subtracted NH images, which could only be obtained during the second run when Lulin was at its brightest. Unlike any of the other species, NH exhibited a single, radial feature that abruptly changed directions from the WSW to the ESE between Feb 24 and 27. This interval exactly surrounds minimum phase angle and the Sun's position angle (PA) changing from 112° to 289°. We can only conclude that we are detecting an ion tail within the NH filter, whose emission dominates over that of NH itself. Having seen no evidence of this feature in other gas filters, we consider it highly unlikely that the source is $CO^+$. Another possibility, $N_2^+$, recently detected in Comet C/Pan-STARRS (2016 R2) (Cochran and McKay 2018), has emission bands close to the NH bandpass, but likely too weak compared to $N_2^+$ bands that should have been evident in the CN filter. Unbeknownst to us, the contaminating ion had been detected decades earlier and identified as the 1-0 band of $OH^+$, with the 0-0 band also identified just longward of 3500 Å, by Festou et al. (1982), and these identifications were later confirmed with ion tail spectroscopy of 1P/Halley by Wychoff & Theobald (1989). Our detection of this water group ion is also consistent with the detection of a faint ion feature in the R-band filter, presumably $H_2O^+$. As to why this is the first time that we've detected it in NH images, it is likely a combination of several factors that make Lulin somewhat unique – having a very small phase angle so that an ion tail remains mostly along the line-of-sight and within the field-of-view, a high gas production rate needed to produce a sufficient number of ions, a depletion in NH (discussed in Section 5), and a sufficiently bright comet for us to even utilize the NH filter.

Due to the limited number of images of gas species other than CN, and the fact that the dust behavior is significantly different than CN, the remaining analyses for this section and most modeling sections are focused on our CN images.

*3.3 Rotation Period*

At the end of our first observing run (Jan 30 – Feb 2), we examined our enhanced CN images to compare the shape, size, orientation, and relative intensity of the jet features to find image pairs with similar morphology. It quickly became apparent that we did not have adequate phase coverage during the run to determine a rotation period, as we had no matching pairs of CN images. By the conclusion of our second observing run (Feb 24 – Mar 2), however, we had matched 15 pairs of images of varying rotational phase, which were obtained on 16 nights (including some "snap shots" obtained between the main runs) from February 18 to March 3 and are listed in Table 3. We were able to derive an apparent, i.e. synodic, rotation period of $42.0 \pm 0.5$ hr with these images (Knight and Schleicher 2009a). With ~10 hours of coverage each night during the second observing run, it was possible for us to rule out the shorter (21 hr, 14 hr, 10.5 hr, etc.) aliases. Biver et al. (2009) later reported a rotation period of $41.04 \pm 0.72$ hr, consistent with our solution to within the uncertainties. They obtained this rotation period from radio and sub-mm spectroscopy taken from mid-February to early March, a time frame encompassed by our interval. As will be discussed in detail in Section 4.5, we were able to extract a precise sidereal rotation period by using our model and extending the time frame to encompass our three main observing runs. We were ultimately able to constrain the sidereal rotation period to 41.45 ± 0.05 hr.

[TABLE 3 here; Apparent rotation periods measured from coma morphology]

### *3.4 CN Jet Measurements and Projected Velocities*

As described in Section 3.1, we observed Lulin's CN corkscrew features to move steadily outward during a night of observations. Measuring the position of each jet on each viable CN image provides us with two important pieces for our comet model – a way to quantitatively compare the morphology of the CN jets from the images with our comet model (discussed in Section 4.3), and the ability to extract the CN projected outflow velocity for each observing run (discussed on Section 4.4). The process we used to obtain the jet measurements is detailed in Schleicher & Woodney (2003) and Farnham & Schleicher (2005), though we briefly describe it here. In short, we divided each enhanced image into thirty-six 10° wedges. We then extracted a radial profile from each wedge, where each profile is the mean brightness from that collapsed wedge. This process greatly improves the signal-to-noise ratio further from the nucleus, and it allows us to measure the distance at which each arc from a corkscrew feature crosses that wedge.

After the brightest location (i.e. arc position) of each jet was recorded for the wedges, we converted the measurements from image pixels to projected distances from the nucleus in km. Using our second imaging run (late Feb through early Mar), where we had the most extensive coverage, we calculated the slopes of the measurements from the successive CN images. Using this method, we extracted a mean projected outflow velocity of ~0.48 km s$^{-1}$ for this time frame. Once we imported the arc measurements into our model, we were able to use them as overlays, and to subsequently use our model to obtain and refine mean outflow velocities during both the preceding and following observing runs. The overlays also served as a check to our model solution, since the measurements needed to match the arc positions in our model. A discussion of this process and our results are in Section 4.3.

### *3.5 Determination of the Pole Orientation*

The two corkscrew jet features apparent in our CN images allowed us to constrain Lulin's rotational axis orientation, as the shape indicates each jet is near a pole (see Figures 1, 2 and 3). Under this assumption, the rotational axis is the midpoint of the cone swept out by each jet. We measured the approximate axis orientation through both of the jets simultaneously, using select CN images from each of our three main observing runs, listed in Table 4. For each run, this then defines the PA of a great circle on which the rotational axis falls. Because there was no evidence of a systematic offset with the measurements, we simply used the mean value per run, as plotted in Figure 6. Although the three curves, each representing the mean position from each run, are roughly parallel, there are two intersections in each hemisphere where two curves cross and the third curve is in close proximity (see Figure 6). The two northern solutions were at a RA/Dec of about 90°/+27° and about 137°/+17°; the southern solutions are simply the opposite pointing poles, thereby having the opposite sense of rotation. Using our model, we were able to differentiate these potential solutions and ultimately determine the appropriate pole solution (see section 4.2).

[Table 4 here: List of P.A.s measured from the corkscrews (obs date, time, angle measured)
[Figure 6 here; pole solutions plot]

## 4. MODELING JET MORPHOLOGY

The image dataset we used for our model consists of multiple nights of observations, covering most rotational phases over a wide range of viewing geometries. The bulk of our images come from the three main observing runs spaced approximately one month apart between 2009 Jan 30 and Apr 1, encompassing the period of +19 to +81 days after perihelion, and we have additional images out to May 15 (+124 days after perihelion; see Figure 7). We removed the average radial profile on all of our images as described in Section 2.3 to more clearly define the prominent features on which we based our models.

[Figure 7 here; Rotational phase coverage of our CN images]

*4.1 The CN Model*

Our Monte Carlo jet model is based on the comet's reference frame and then undergoes a series of coordinate transformations to match what was observed from Earth at any particular time. It has a large number of parameters, including the tilt and orientation of the rotation axis, the location and size of multiple source regions, the outflow velocity, the probability that a particle is emitted from the surface based on the amount of solar radiation that in turn varies with rotation and season, and, for dust grains, the effects of radiation pressure. Additional details were presented in Schleicher & Woodney (2003) for our modeling of Comet Hyakutake (1996 B2), and in Schleicher et al. (2003) in our analyses of the strong seasonal asymmetries in Comet 19P/Borrelly. A modified version of the model was used to successfully match the ejection plume from the Deep Impact encounter with Comet 9P/Tempel 1 (Schleicher et al. 2006). The major differences with modeling CN jets here as compared to dust are that gas molecules are in some ways easier – with all particles having the same mass, there is much less dispersion in outflow velocities and only very small radiation pressure effects – and in other ways more complicated – including a greater dispersion in the initial direction of molecules as they leave the surface of the nucleus, the effects of lifetimes for the parent and daughter molecules, and an additional vectorial motion of the observed species caused by the excess energy of dissociation of the parent molecules. In either case we follow a top down, iterative approach, using reasonable starting values for many parameters while solving others, then adjusting those assumed values. For Lulin, having a preliminary rotation period (see Section 3.3) and only a few viable pole solutions (see Section 3.5) greatly simplified the process, while having two jet features originating from the same body also helped test our solutions.

*4.2 Pole Solution*

Our first modeling task was to examine the four potential pole solutions, discussed in Section 3.5, and determine the correct solution. Recall that we measured the PAs of the projected rotation axis, which provided a very strong constraint perpendicular to the ecliptic but less well constrained along the comet's orbit. The model was set up with each potential pole solution in turn, with a jet in place near each pole to approximate the appearance of our CN images. We then visually compared the positions and directions of both of the CN corkscrew features in the images and in the model during each observing run. The "southern" pair of solutions were eliminated because their sense of rotation was opposite of that of the observed corkscrew. The northern solution having the larger RA, at RA/Dec ~ 137°/+17°, resulted in the jets passing in front of one another and swapping sides during the observing interval, was also in direct conflict with our imaging; using the "snapshots" obtained between the main observing runs we were able to confirm that the two jets never crossed during this time. In fact, even the remaining solution, at RA/Dec ~ 90°/+27°, reached an overlap of the jets late in the apparition, and we had to move the pole further, to 81°/+29°, for the optimum solution though this final answer was only reached after iterating the various other model parameters described in the next sub-sections. Note that this final solution, also marked on Figure 6, lies within the region bracketed by the three great circle solutions from the three runs, and corresponds to a pole obliquity of 97° and orbital longitude of 120° in the comet and model's reference frame.

*4.3 Preliminary Source Locations and Sizes*

Once the correct pole solution was determined, we proceeded to use our Monte Carlo jet model to more accurately determine the locations of Lulin's two CN jets. To determine their locations, we first examined our best images from our second observing run, when the comet was at its closest approach to Earth, from Feb 24 through Mar 2 *($\Delta T$ = +45 to +52 day)*. Using our synodic 42.0 hr rotation period, we chose ten images spaced approximately 0.1 apart in rotational phase from across the observing run. Then, moving to our model, we assigned reasonable values to the main parameters, such as outflow velocity, since these have little effect on the overall shape of the spirals (we later refined the model parameters once we determined a viable position for each active region).

We next performed a grid-pattern search for each jet separately to narrow down the approximate latitude for each active region. This is an iterative process that was first done on a coarse scale and then refined to a more narrow, targeted search. We matched the locations, shapes, and relative brightness of the CN jets to the images, while paying attention to the solar altitude to understand the availability of sunlight on the active regions. We added an overlay of arc wedge measurements for each image to the model (see Section 3.4) to quantitatively match the model to the measured positions of the jets. Once we narrowed the range of viable source locations, we conducted a similar, iterative, grid pattern search for these narrowed ranges for both the earlier run (Jan 30 through Feb 2, $\Delta T$ = +19.8 to +21.9 day), and the later run (Mar 28 through Apr 1, $\Delta T$ = +76.5 to +80.6 day). Through this process, we were able to determine a model solution that endured throughout the entire span of our observing runs except in one respect – the source longitudes systematically shifted from run to run, directly implying that the rotation period would need adjusting. At this stage of the process, our approximate source location for the east jet required a latitude of about -80° while the western source needed to be near +75°, and the offset between the source longitudes was between 110° and 140°. Finally, additional grid-pattern searches were conducted to narrow down the source sizes that best reproduced the appearance of each jet, and these yielded radii of 8°-15° and 15°-25°, respectively.

*4.4 CN Outflow Velocity*

Early on in the modeling process it became apparent that the jets' motions were essentially in the plane of the sky during the second run, and therefore the measured, i.e. projected velocities derived from the arc wedge measurements (Section 3.4) should be close to the actual outflow velocities. We therefore directly used the measurements to provide a functional form for the acceleration and final coasting velocity for the gas in our initial iterations of the model. This had a starting velocity at the nucleus of 0.10 km s$^{-1}$, reached a coasting velocity of 0.48 km s$^{-1}$ at 20,000 km, and had a simple parabola with these constraints during the acceleration regime. We found that the same shape also worked well during the first and third run, only requiring the application of scaling factors due to the differing distance of Lulin from the Sun. By later iterations in the modeling, it became apparent that the best overall solutions were obtained with systematically lower velocities throughout the apparition, but again with the same basic shape for the function. This is most likely due to a combination of issues, including projection, viewing, and distance changes within runs. Our final terminal velocities were ~0.62 km s$^{-1}$ at ~26,000 km for run 1, ~0.40 km s$^{-1}$ at ~17,000 km for run 2, and ~0.32 km s$^{-1}$ at ~13,000 km for run 3, where the last value is much less certain because of the poor S/N, and values for both the first and third runs are unusually dependent on the other parameters such as the pole solution and source latitudes because of the very large projection effects at these times.

*4.5 The Sidereal Rotation Period*

Similar to how an apparent lightcurve period differs somewhat from the underlying sidereal period of a body because of changing viewing geometry, so too does the time between morphological features matching locations in the sky plane differ from the sidereal period because of changing projection effects. Moreover, while determining a source's latitude is relatively easy, a source's longitude is highly dependent on the velocity relation applied when tracing a jet back to its origin, hence the need to finalize the velocity behavior prior to solving for the fundamental period. To determine the sidereal period, our goal here is to remove any apparent drift in the derived longitude of a source region from one run to the next, a direct byproduct of having the incorrect period. It became evident that we needed a somewhat shorter period than our initial synodic value of 42.0 hr, and our final solution is 41.45 ± 0.05 hr (1.727 ± 0.002 day), though the actual uncertainties are probably double the quoted values due to their interdependence on the specific velocity relations we adopted. Note that we did not detect any evidence for a change in Lulin's sidereal period during the apparition; the measured synodic period was essentially unchanged for the duration, a single solution for the sidereal period and the source longitudes worked throughout, and one would expect minimum torqueing in the rotational direction from source regions located very close to the poles.

*4.6 Refined Source Latitudes, Longitudes, and Radii*

Having fine-tuned axis orientation, the outflow velocities, and the sidereal period, we performed a final refinement to the solutions for the source region latitudes, longitudes, and radii, again first fitting the second run's images, then confirming and making small adjustments based on the other runs. Having set the relative proportion of molecules available for release from the two source regions in proportion to the source areas (i.e., the same flux per unit area), we also attempted to match relative brightness both along the jets and from jet-to-jet as a function of rotation and season. Of course the actual rate of release also depends on the amount of solar radiation available, another component of our investigation that will be discussed in Section 4.9.

We found that the east jet fits best at a latitude of about -80° while the best position for the west jet is at +77°. The difference in longitude between the two jets is approximately 120° and, based on a sidereal period of 41.45 hr, the longitudes are about 125° and 245°, respectively, where 0° longitude is set as the anti-sunward location at perihelion. The east jet radius is around 10°, and the west jet is larger with a source radius near 20°, and thus 4× the active area. The strength of the west jet throughout the apparition, even when the solar altitude is low, supports our conclusion that its source region is very likely physically larger than the source region for the east jet, and the larger size is a better match for the detailed jet morphology.

*4.7 Summary Model Solution and Comparison to CN Jets*

Given that our best solution is the result of a series of compromises along with some assumptions we know are unrealistic, in particular a spherical nucleus with circular source regions, we think it holds up remarkably well throughout the apparition. The peak in brightness for the images is consistent with a thermal lag, with peak production occurring a few hours after the Sun's crossing of the meridian, a process we are unable to model at this time. The final results for our CN model solution are summarized in Table 5, and in Figures 8, 9, and 10, we compare representative images to our best solution for the jet model. The top-down view of the comet from the model is also included in these figures to more easily visualize the 3-D positions of the jets and the degree of the projection effects for each time step.

[Table 5 here; summary model results]
[Figures 8, 9, 10 here; Images and model solutions, runs 1, 2, 3]

*4.8 The Dust Model*

Near the end of Section 3.1, we described the various dust features seen with the *R*-band, some of which are thought to be either young and/or old tail material, emitted days or even months before a particular image, and still within the field of view due to a combination of very low velocities coupled with large projection effects. For our modeling here, we disregard these likely tail features and only focus on the other, inner coma structures. These are not coincident with the CN jets, though there is a somewhat broader or more diffuse feature towards the west during the first run and towards the east by the third run. We identify these diffuse features as dust indeed emitted sequentially by our two CN sources, but only one is evident each month because gas flow must be near a maximum, i.e. when the Sun is high in the sky, for there to be sufficient flow to lift appreciable quantities of dust. As further discussed in the next subsection, this only takes place in the west early in the apparition and in the east late in the apparition. The lack of any rotational signature in these west and then east dust features is simply due to a combination of low dust velocities ($\leq 0.1$ km s$^{-1}$) – causing overlap of successive rotational cycles – along with a dispersion in dust velocities associated with a range of particle sizes – further smearing any detailed structures.

The remaining, closest in features are the most distinct. As also noted in Section 3.1, a feature is evident towards the east, southeast, or south (at differing rotational phases) and curls towards the south and then west during the first run, ultimately overlapping with the western feature. By the time of the second run, this feature has disappeared, while two nearly linear features are evident towards the SSW and NNE, alternating in brightness with rotational phase. Extensive testing using our two CN jets model solution while varying outflow velocities, radiation pressure effects associated with particle sizes, and other differences between dust grains and CN molecules did *not* yield any explanation for these dust features in the inner-most coma. Failing this, we then explored the possibility of a third source region.

Given the known viewing orientations it was apparent that a single source near the comet's equatorial region might yield suitable coma structures, and we indeed found a solution that reproduced the appearances in both runs. This source produced a jet that not only matched the changing initial position angle of emission and the hooked shape in the first run, but also yielded both linear features in the second run because the rotational axis was then very near the sky plane and the equatorial jet was seen nearly edge-on from Earth. This third source region has a latitude of about -10°, a longitude of about 150°, and a highly uncertain source radius of ~10°. Other properties are consistent with what we found for dust emitted from the two gas source regions: a mean velocity of ~0.1 km s$^{-1}$ during the first run, ~0.09 km s$^{-1}$ during the seond run and ~0.07 km s$^{-1}$ during the third run; a significant dispersion in outflow velocities presumably associated with the range of particle sizes; and a dominant particle size of ~2 microns along with a large dispersion in the particle sizes as expected for a particle size distribution of dust, with this size based on the observed radiation pressure effects. Similar to the CN jet modeling, we conclude that the imaging is most consistent with a thermal lag, where peak production takes place a few hours following the Sun's crossing of the meridian. However, the dust jet is only active when the Sun is high in the sky, when the gas flow is at a maximum.

[Figures 11 and 12 here; model comparison with dust images]

The mystery, of course, is what gas is lifting the grains from this third, equatorial source and why aren't the normal minor species present? One possibility is pure water ice. This would also be consistent with the finding from Section 3.2 that OH distribution did not match that of CN, $C_2$, or $C_3$, and that we don't even see this third jet in images for the carbon-bearing species.

*4.9 Explanation of Diurnal and Seasonal Effects*

The corkscrew CN jets clearly exhibited brightness variations that showed diurnal variations. As always, we started with the obvious cosine function of the Sun's angle from the zenith when computing the amount of available solar radiation driving the ice sublimation and proportion of gas molecules leaving a source region. This worked well except in two respects. First, there is strong evidence for a thermal lag such that peak production was in the early afternoon rather than when the Sun crossed the meridian, an effect we are not currently able to model. Second, gas continues to be emitted at a low rate even at night, at about 5% of peak production with the Sun overhead, though this also shows evidence of a decline during the night, with the minimum rate of release occurring near dawn, again consistent with a thermal lag. Similar behavior has been seen in some other comets; we assume that this is simply due to the ice remaining warm from thermal lags and prior solar heating. The dust, however, behaves quite differently from the CN. Not only is there no dust lifted from the nucleus at night, but essentially none is released when the Sun is low in the sky. In fact, our best match was assuming an extremely steep functional form from the zenith, cosine to the 6th or even 8th power, i.e. the Sun needed to be within about 30° of the zenith to get sufficient gas flow to lift the dust grains.

Based on these diurnal patterns, the observed seasonal behavior is quite understandable when one examines the sub-solar latitude on Lulin's nucleus as a function of time. As we show in Figure 13, the sub-solar latitude is at +10° at the start of the first imaging run ($\Delta T$ = +20 day) and drops smoothly to -35° by the end of the third run ($\Delta T$ = +90 day). Thus the western jet dominates during the first run, having somewhat more intense sunlight on the source region combined with a much larger surface area, while by the third run this source will have reached "perpetual winter" but continues to release gas at a low level. This readily explains why the eastern jet dominates by this time, but more surprising is that the eastern jet is on at all during the first run. The Sun would have only just appeared above the horizon for the first time since more than 20 months earlier, when the Sun was last as far south as +20° and thus the 10° radius source centered at -80° latitude would have been illuminated. Moreover, at that time Lulin had a heliocentric distance of nearly 7 AU, so we assume there is no residual heat causing activity. Rather, a very low sun angle suggests that a more volatile species such as CO or $CO_2$ is the driver, rather than water ice. Note that our definition of latitude is based on the normal angle to the surface and not the geocentric (or, i.e., cometocentric) definition; thus any departure of the nucleus' true shape from our nominal sphere is largely irrelevant to this discussion of sun angles.

[Figure 13 here; sub-solar and sub-Earth latitudes]

We also see from Figure 13 that the peak in the sub-solar latitude took place 10 weeks prior to perihelion at a value of +83°, with this latitude directly associated with the 97° obliquity of the axis. The most negative sub-solar latitude, at -83°, occurs just over a year following perihelion, at a distance of 3.8 AU. To better understand some of the geometries involved during the apparition, also plotted is the sub-Earth latitude as a function of time. The reversal is of course near the time of the close passage by Earth, with the Earth crossing the equator on Feb 23, and our view being near pole-on at the start of our first run (+60°) and the opposite pole at the end of our third run (-72°), hence the corkscrews being seen at large projections.

## 5. PHOTOMETRY ANALYSES AND RESULTS

### *5.1 The Photometry Data Set*

We now turn to the results from our narrowband photometric measurements; recall from Section 2.2 that we have very limited photometry data due to having several runs weathered out early in the apparition and our focus on imaging when the comet was bright enough to do so. The reduced fluxes and associated aperture abundances (as logarithms) are listed in Table 6, while the resulting log production rates, *Q*, and *A(θ)fρ* values are given in Table 7 and plotted as a function of log $r_H$ in Figure 14 where filled symbols are used for data on the sole pre-perihelion night, and open symbols after perihelion. We include only the upper or "+" log sigmas in Table 6 since the uncertainties are unbalanced in log space; the lower "-" values can be calculated. The vectorial-equivalent water production value is included in the last column of Table 7. The measurements for most species are self-consistent within the photometric uncertainties for each night of observations; the NH measurements are an exception, with more spread in the data and higher uncertainties.

[TABLE 6 here; Fluxes and Aperture Abundances]
[TABLE 7 here; Photometric Production Rates]
[Figure 14 here; Photometry 6-panel]

### *5.2 Composition and Behavior with Heliocentric Distance*

Lulin's mean production rate ratios for the gas species, as well as for dust-to-gas, place the comet in our typical compositional class (cf. A'Hearn et al. 1995; Schleicher & Bair 2016). Very small trends with aperture size were evident on some nights of observations, but since these can be directly attributed to the rotational variability of the comet there is no need for an adjustment in our analyses.

The outbound measurements follow a smooth trend of decreasing production rates, and the OH and CN log *Q* vs log $r_H$ slopes are the shallowest, and essentially identical, at -2.75±0.2 and -2.77±0.1, respectively. The slope for $C_3$ (when excluding the conspicuously low point during the last night) is -3.18±0.1, while the $C_2$ slope is steeper at -3.60±0.1. The NH slope is by far the steepest of the gas species with a value of -9.64±0.5. The higher photometric uncertainties on the larger $r_H$ points are insufficient to explain the discrepancy, leaving two other possible causes. First, is that there is a large compositional change in the NH abundances between the two source regions, with the mixing ratio for NH much lower in the eastern source that dominates gas production later in the apparition. However, the NH/OH production rate ratios were quite similar very late in the apparition to those from our first, pre-perihelion night (and at a similar $r_H$) and since our modeling indicates a switch in which source region dominates emission at these two times, a compositional difference cannot be the cause.

Our second possible explanation for the very steep $r_H$-dependence appears much more likely, and is based on the imaging of NH in late February that suggests the presence of ion contamination. Whatever ionic species this would be – the most likely candidate being $OH^+$ – it is easy to imagine that such contamination would be much greater at smaller $r_H$ when the production of ions is highest and/or at small phase angles when the ions would remain along our line-of-sight and therefore within the photometric aperture for longer periods of time. Also, the water production rate in late-February was about $5 \times 10^{28}$ molecules s$^{-1}$, coincidently near the apparent threshold for ion production in 1P/Halley (Schleicher et al. 2015), thereby suggesting that ion emission was no longer a significant contaminant late in Lulin's apparition. The lack of obvious NH jet morphology in the images suggests that the ion emission might even dominate the signal in the NH filter (though the narrow ion tail as compared to the far broader distribution of neutral species makes a proper comparison difficult). Additional evidence comes from the NH/OH ratio from month-to-month. While the mean value places Lulin into the typical category, at larger $r_H$ the NH/OH value would place the comet into the NH depleted compositional group (Schleicher & Bair 2016). Similar behavior was reported for several comets by Festou et al. (1982; and references therein), who in fact noted "…in the form of a peculiar behavior of the heliocentric variation in the NH/OH intensity ratio."

Dust, after excluding the distinctly low point during the last night of observations, has a slope of -3.88, but this is artificially steep due to phase angle effects, and normalizing for phase angle, as discussed later, yields -1.74±0.1 – much more shallow than for any of the gases. For comets receding from the Sun such a shallow slope is common, and we have often attributed it to a significant component of heavier, much slower moving grains remaining within our field of view that were lifted off of the surface during maximum gas flow near perihelion.

Even though we have only one night of data prior to perihelion, and at a larger distance than post-perihelion measurements, we can look for pre-/post-perihelion asymmetries by simply extrapolating our fits of the post-perihelion data. While the uncertainties of some gas species are higher due to faintness and distance, in all cases it is evident that Lulin was more active before perihelion than after at 2.6 AU from the Sun. Since Lulin's west jet source is intrinsically larger, and our model indicates it was exposed to the Sun while the comet was inbound, this higher activity before perihelion is as expected.

### 5.3 Water Production, Effective Active Area, and Model Comparison

Our highest water production rate of approximately $6.30 \times 10^{28}$ molecules s$^{-1}$ was measured on 2009 Feb 26 ($\Delta T$ = +46.7 day), our first night of photometry after perihelion. However, we can use our fit to the OH data to extrapolate back to perihelion and estimate the peak water production at that time, with a result of $1.0 \times 10^{29}$ molecules s$^{-1}$. This value implies a maximum effective active area of 42 km$^2$ using a water vaporization model based on the work of Cowan and A'Hearn (1979; see also A'Hearn et al. 1995 but a factor of two error has now been corrected – M. A'Hearn 2010, private communication). Collectively our water production rates for Lulin, listed in the last column of Table 5.2, indicate a median effective active area of 38 km$^2$; this number is well above the median value for active area for all of the dynamically new comets in our database, which is 24 km$^2$. Assuming the case that the entire surface of the nucleus is active, this would imply a lower limit for the radius of ~1.7 km. We know from our images, however, that the entire surface of the nucleus is not active and that most activity is driven by two main source regions. Using our model, we were able to derive an estimate of the radius of each source region in relation to the size of the nucleus, and our primary (i.e. western) source is only about 3% of the nucleus. If one includes the eastern source and the dust (equatorial) source, one reaches a total active fraction of only 4-5% of the nucleus. In turn, this implies a nucleus radius of perhaps 8 km. However, the true value may be significantly smaller if much of the measured water originates either from an additional, isotropic source such as leakage over the entire surface, or from ice grains in the coma rather than water molecules vaporizing off of the surface.

Several other investigators also measured Lulin's water production rate but inter-comparisons are made difficult for several reasons. First, nearly all of the published measurements were obtained near the end of January, nearly a month prior to our own earliest post-perihelion observations. Extrapolating our own results back to the end of January, using the post-perihelion $r_H$ dependence, yields a $Q(H_2O)$ of $9 \times 10^{28}$ molecule s$^{-1}$. This is about 25% more than Bodewits et al. (2011) determined using Swift, but less than the values reported by Gibb et al. (2012) using NIRSPEC on Keck, with results ranging from $1.1-2.5 \times 10^{29}$ molecule s$^{-1}$ within a three-day interval and that they attribute to rotational variability. The most extensive, but unpublished, set of water determinations were using SOHO/SWAN H-✓ measurements (Combi, personal comm.), and these are only 15% higher than our own extrapolated value at the end of January. However, they show a steeper $r_H$-dependence during January than our constant slope of -3.25 (water is steeper than OH by -0.5), and a nearly level slope during six weeks from early March to late April.

### 5.4 Dust behavior

The dust in Lulin's coma exhibits several atypical characteristics when compared to most other comets. First, there is almost no trend in $A(\theta)f\rho$ with aperture size, implying that the radial profile of dust essentially follows the canonical $1/\rho$ fall-off, unlike the majority of comets who exhibit a much steeper profile. Even more unusual is that the dust exhibits, to within the uncertainties, no reddening at shorter wavelengths even in the ultraviolet, again unlike most comets. Specifically, Δ log $A(\theta)f\rho$ for green – blue is 0.00, green – UV is -0.02, and blue – UV is -0.01. However, $A(\theta)f\rho$ in the red is systematically higher by about 44% as compared to bluer wavelengths (see the blue and red values from the narrowband images given in Table 8). This characteristic is also abnormal, and implies unusual characteristics for the dust grains, possibly associated with the compact aggregates claimed by Woodward et al. (2011).

While we routinely normalize the $A(\theta)f\rho$ values for phase angle effects using our now-standard composite phase function (Schleicher & Bair 2011), our dense imaging coverage surrounding opposition presented a rare opportunity to investigate in detail phase effects at small phase angles. We therefore extracted *R*-band fluxes and, using on-chip stars for calibration, determined relative $A(\theta)f\rho$ values throughout the second observing run, during which the phase angle decreased from 9° down to 0.7° and back up to 17°. The results are shown in the top panel of Figure 15, along with one night from the first run, after removing the nominal $r_H$-dependence of -1.74 that we previously found. While there is some evidence that Lulin's phase curve is slightly shallower than our standard curve, they are extremely similar, in particular at very small phase angles.

[Figure 15 here; Afrho items, two panels]

In the bottom panel of Figure 15 we again plot dust production as a function of the log of the heliocentric distance (see Figure 14), but now normalized to 0° phase angle using the just-confirmed phase function from Schleicher and Bair (2011). These results for the green continuum, now on a nearly straight line, are overlaid with the blue continuum values from the narrowband imaging; no adjustments were made for color as we previously found essentially no difference between the green and blue results. As mentioned in Section 5.2, the $r_H$-dependence for the dust after perihelion, with a slope of -1.74, is much more shallow than that of any of the gas species, though both gases and dust do have a similar pre- vs. post-perihelion asymmetry, with higher production rates before perihelion compared to the extrapolated value at the same heliocentric distance using the slope measured for dust. Finally, the mean dust-to-gas ratio, based on the phase adjusted $A(\theta)f\rho$ to Q(OH), was -25.34, near the middle for both dynamically new comets and comets over all.

## 6. DISCUSSION AND SUMMARY

Our enhanced CN images reveal that Lulin has a double corkscrew morphology, with two near-polar jets oriented approximately E-W. Our large dataset reveals seasonal changes in bulk CN and dust morphology, and nightly rotational variability through the CN and dust filters. We acquired images in additional narrowband filters when observing circumstances were favorable, allowing us to compare the behaviors of multiple gas species during the main observing runs. As expected based on previously observed comets, $C_3$ features exhibited shapes quite similar to, but not extending as far as, CN. The normally broader jets expected for $C_2$ because of its multi-generation parentage appeared more confined early in the apparition for Lulin, possibly suggesting an extended collision zone encompassing the region where most of the $C_2$ parent dissociations take place. The morphologies observed through the *R*-band (dust), OH, and NH filters were all quite different from the carbon-bearing species.

*CN Model Solutions*
We measured the position angle for Lulin's rotational axis from multiple CN images, which we then used to determine its potential pole solutions. We needed to turn to our model, however, to precisely pin down the axis location. Since Lulin was traveling in Earth's orbital plane and we never saw the pole cross our plane of view (i.e. the two CN jets always stayed on their respective sides of the nucleus), the model allowed us to pin down a pole solution that fit within the potential ranges and matched our images, without allowing the source regions to flip sides and preserving the bulk seasonal changes seen in our images. We determined that Lulin's pole has an obliquity of about 97° and an orbital longitude of the pole of around 120°, corresponding to RA/Dec = 81°/+29°. This pole solution is secure to within a few degrees, a result of the circumstances above combined with our extended intervals of observations.

With a pole solution in place we replicated, through a highly iterative process, the positions of both CN corkscrew features and we constrained the approximate radius of each source region. The east source area has a lat/long of -80°/125° and radius of around 10°, while the western source area is at lat/long +77°/245° and has a radius near 20°, approximately 4× larger than the eastern source. We emphasize that our final model answer is a compromise solution, as slight adjustments to the parameters could help during certain time frames but then cause the model to not match the images during later and/or earlier time frames. Comet shape and topography are the most likely explanations for this, as our model assumes a spherical comet and circular source regions, but thermal phase lags are also a contributing factor.

When we combined the model constraints with our long timeline of images, we derived a highly refined sidereal rotation period of 41.45 ± 0.05 hr, a solution that held up for the duration of our three main imaging runs with no evidence for a need to either slow down or speed up the rotation period as time progressed. We found Lulin's CN outflow velocities to steadily decrease as the comet moved further from the Sun. The final terminal velocity was ~0.62 km s$^{-1}$ at ~26,000 km during our first run, and decreased to ~0.32 km s$^{-1}$ at ~13,000 km during our third run. The eastern jet is active with very little sunlight during our first run, suggesting a more volatile driver such as CO or $CO_2$ is necessary, and the lack of OH jets coinciding with the carbon-bearing jets supports this.

*Dust Results*
Our enhanced *R*-band images revealed various dust features, including dust tails and inner coma structures associated with activity closer to the nucleus. Our dust model focused on the inner coma structures, and we realized a third source region near the comet's equatorial region was necessary to reproduce these features. This additional source is located at a latitude of ~10°, a longitude of about 150°, with a highly uncertain source radius of around 10°. The model fit required a large range in particle sizes, with dominant sizes ~2 microns, and with dust activity only when the Sun is high in the sky, i.e. when gas flow is at a maximum. Our conclusions are consistent with results from Woodward et al. (2011) and Joshi et al. (2011), who also determined the dust in Lulin is dominated by large particles. Mean dust velocities also decreased with distance from the Sun, going from 0.10 km s$^{-1}$ during the first run, to 0.09 km s$^{-1}$ during the second run, and to 0.07 km s$^{-1}$ during the third run. Due to Lulin's rapid passage through extremely small phase angles, we investigated its phase angle effects and compared them to our standard solution based on Halley and found them to be extremely similar, particularly at small phase angles.

*Water Results*

Our water production rates from photometry indicate a median effective active area of 38 km$^2$, implying a lower limit of the nucleus radius of around 1.7 km. Using our source radius estimates from the model, we calculated an approximate active fraction of the nucleus of only 4-5%, corresponding to a nucleus radius of ~8 km. The radius may be significantly smaller, however, if much of the measured water originates either from an additional, isotropic source such as leakage over the entire surface, or from ice grains in the coma rather than water molecules vaporizing off of the surface. The need for an additional source region in our dust model to explain some of the observed features supports the idea that additional areas of the nucleus are active, and it likely explains the distribution of OH in our images, which is much more uniform than what we see for the carbon-bearing species.

*Ions*

While not looking for ions, we saw clear evidence for ion tails in both the *R*-band filter, where $H_2O^+$ is the obvious species, and in the NH filter, where we had not previously detected an ionic feature but later learned that $OH^+$ has its second strongest emission band; both species are also water group ions. This detection within the NH filter was likely aided by a number of factors including the strong projection effects at opposition with our line-of-sight down the tail, and the overall high production rates. As a contaminant, this can also explain the extremely steep $r_H$-dependence seen for NH, and its apparent change from being NH depleted at large $r_H$ as compared to "typical" composition at smaller $r_H$, as one would expect significant ionization only closer to the Sun.

*NH Depletion*

When looking only at its mean NH/OH ratio, Lulin is in our typical compositional class along with the majority of the dynamically new comets in our database (Schleicher & Bair 2016). A smaller portion of dynamically new comets, about one-fifth of the ones we have measured, have mean production rate ratios that place them in our newly defined NH depleted class, a group including comets with lower NH abundances with respect to water, but with no depletion in either $C_2$ or $C_3$. We found one other comet (C/LINEAR 2002 T7) classified as dynamically new and typical in our database that behaves similarly to Lulin, changing NH composition with $r_H$. This implies we have potentially mis-classified at least one, and possibly two NH depleted comets as typical in our database due to $OH^+$ ion contamination, and suggests there may be an even higher number of NH depleted comets.

*Is Lulin Really Dynamically New?*

Although Comet Lulin is classified as a dynamically new comet, having a $1/a_0$ value of $31 \times 10^{-6}$ $AU^{-1}$, and therefore smaller than either common definition, $< 100 \times 10^{-6}$ $AU^{-1}$ or $< 50 \times 10^{-6}$ $AU^{-1}$, it shares none of the common attributes of a comet arriving into the inner solar system for its first time. In particular, it showed no evidence of hypervolatile activity typical for first-timers but rather appeared nearly asteroidal at discovery near 7 AU with only a small coma. It then brightened like a normal long-period comet as it approached the Sun, rather than having a near-constant brightness due to the loss of hypervolatiles (cf. A'Hearn et al. 1995). The simple existence of jets from isolated source regions directly implies that much of the surface is relatively inert - itself an evolutionary process - indicating it has approached the Sun perhaps many times in the past. Finally, and more indirectly, Woodward et al. (2011) conclude that the dust in Lulin is dominated by large and compact aggregate particles that are typical for old period comets, while new comets tend to have more porous aggregates. We, therefore, conclude that it is highly likely that Lulin, while being dynamically new in the statistical sense of the definition, behaves like a long-period comet and "has passed this way before." Anecdotal evidence suggests a greater fraction of "new" comets exhibit "old" behavior than would be expected based on statistical arguments, and that the dividing line between dynamically new and old should be re-investigated based on physical properties exhibited by the comets themselves.

## ACKNOWLEDGEMENTS

We thank Len Bright, Georgi Mandushev, Brian Skiff, and Alan Gersch for the acquisition of the snapshot images. The jet model was created within the Data Desk software package. This research has made use of JPL Horizons (Giorgini et al. 1996) and SAOImage DS9, developed by the Smithsonian Astrophysical Observatory. We gratefully acknowledge support from NASA's Planetary Astronomy Program and Planetary Atmospheres Program.## REFERENCES

A'Hearn, M. F., Millis, R. L., Schleicher, D. G., Osip, D. J., & Birch, P. V. 1995, Icarus, 118, 223
Ahn, C. P., Alexandroff, R., Prieto, C. A., et al. 2012, ApJS, 203, 2
A'Hearn, M. F., Schleicher, D. G., Feldman, P. D., Millis, R. L., & Thompson, D. T. 1984, AJ, 89, 579
Bair, A. N. & Schleicher, D. G. 2016, in AAS/DPS Meeting 48 Abstracts, 217.10
Biver, N., Bockelee-Morvan, D., Colom, P., et al. 2009, in AAS/DPS Meeting 41 Abstracts, 23.0522

Table 1. Photometry Observing Circumstances and Fluorescence Efficiencies for Comet C/Lulin (2007 N3)

| UT Date | $\Delta T$ (day) | $r_H$ (AU) | $\Delta$ (AU) | Phase Angle (°) | Phase Adj. log $A(0°)f\rho$ [a] | $\dot{r}_H$ (km s$^{-1}$) | log $L/N$ [b] (erg s$^{-1}$ molecule$^{-1}$) OH | NH | CN |
|---|---|---|---|---|---|---|---|---|---|
| 2008 Jul 31.24 | –163.40 | 2.613 | 1.637 | 7.7  | +0.13 | –19.1 | –15.650 | –14.032 | –13.258 |
| 2009 Feb 26.35 | +46.71  | 1.410 | 0.420 | 0.1  | +0.00 | +13.3 | –14.567 | –13.410 | –12.676 |
| 2009 Mar 21.28 | +69.64  | 1.609 | 1.036 | 36.8 | +0.43 | +16.5 | –14.674 | –13.554 | –12.790 |
| 2009 Apr 21.18 | +100.54 | 1.924 | 2.081 | 28.7 | +0.37 | +18.5 | –14.879 | –13.714 | –12.983 |
| 2009 May 12.19 | +121.55 | 2.151 | 2.704 | 20.2 | +0.29 | +19.0 | –14.975 | –13.810 | –13.093 |
| 2009 May 13.18 | +122.54 | 2.162 | 2.731 | 19.8 | +0.29 | +19.0 | –14.979 | –13.815 | –13.097 |

[a] Adjustment to 0° solar phase angle to $A(\theta)f\rho$ values based on assumed phase function (see text)
[b] Fluorescence efficiencies are for $r_H = 1$ AU, and are scaled by $r_H^{-2}$ in the reductions.

Table 2. CCD observing circumstances for Comet Lulin (C/2007 N3) in 2009.[a]

| UT Date | UT Range | $\Delta T$ [b] (day) | $r_H$ [c] (AU) | $\Delta$ [d] (AU) | $\alpha$ [e] (°) | Elong.[f] (°) | Sun P.A.[g] (°) | Mode | Filters | Weather |
|---|---|---|---|---|---|---|---|---|---|---|
| Jan 30 | 10:29–13:04 | +19.9 | 1.251 | 0.988 | 50.5 | 78.7 | 105.1 | Imaging | R,CN,BC,C3,C2,GC,RC | Photometric |
| Jan 31 | 10:27–13:06 | +20.9 | 1.255 | 0.957 | 50.7 | 80.5 | 105.3 | Imaging | R,CN,BC,C3,C2,GC,RC | Photometric |
| Feb 1 | 9:59–12:49 | +21.8 | 1.259 | 0.926 | 50.9 | 82.3 | 105.5 | Imaging | R,CN,BC,C3,C2,GC,RC | Some cirrus |
| Feb 2 | 9:56–13:11 | +22.8 | 1.263 | 0.895 | 50.9 | 84.2 | 105.8 | Imaging | R,CN,BC,C3,C2,GC,RC | Photometric |
| Feb 18 | 7:56– 8:22 | +38.7 | 1.352 | 0.466 | 32.0 | 133.5 | 112.2 | Snapshot | R,CN | Clouds |
| Feb 19 | 7:28– 7:44 | +39.7 | 1.359 | 0.450 | 28.8 | 138.5 | 112.5 | Snapshot | R,CN | Photometric |
| Feb 20 | 7:52–10:52 | +40.8 | 1.366 | 0.435 | 25.0 | 144.3 | 112.8 | Snapshot | R,CN | Photometric |
| Feb 21 | 10:49–11:09 | +41.8 | 1.374 | 0.424 | 20.8 | 150.4 | 112.9 | Snapshot | V,CN | Clouds |
| Feb 22 | 8:24– 8:55 | +42.5 | 1.379 | 0.418 | 17.1 | 154.5 | 113.0 | Snapshot | R,CN | Clouds |
| Feb 24 | 4:53–13:19 | +44.7 | 1.395 | 0.411 | 8.4 | 168.1 | 112.6 | Imaging | R,OH,NH,UC,CN,C3,BC,C2,GC,RC | Some cirrus |
| Feb 25 | 3:19–13:00 | +45.7 | 1.402 | 0.414 | 4.2 | 174.0 | 112.6 | Imaging | R,OH,UC,CN,C3,BC,C2,GC,RC | Cirrus |
| Feb 26 | 2:57– 4:24 | +46.5 | 1.408 | 0.418 | 0.7 | 179.0 | 115.6 | Imaging | R,CN,BC | Photometric |
| Feb 27 | 2:19–12:00 | +47.7 | 1.417 | 0.428 | 4.0 | 174.3 | 289.6 | Imaging | R,OH,NH,UC,CN,C3,BC,C2,GC,RC | Cirrus |
| Feb 28 | 2:13–12:15 | +48.7 | 1.425 | 0.440 | 7.9 | 168.6 | 289.2 | Imaging | R,OH,NH,UC,CN,C3,BC,C2,GC,RC | Some cirrus |
| Mar 1 | 2:48–11:55 | +49.7 | 1.433 | 0.455 | 11.5 | 163.2 | 288.4 | Imaging | R,OH,NH,CN,C3,BC,C2,GC,RC | Clouds |
| Mar 2 | 2:05–11:33 | +50.6 | 1.441 | 0.473 | 14.7 | 158.3 | 287.5 | Imaging | R,OH,NH,CN,C3,BC,C2,GC,RC | Cirrus |
| Mar 3 | 2:07– 2:39 | +51.5 | 1.447 | 0.489 | 17.2 | 154.4 | 286.8 | Snapshot | R,CN | Clouds |
| Mar 18 | 2:24– 2:44 | +66.5 | 1.579 | 0.928 | 36.2 | 110.3 | 277.3 | Snapshot | R,CN | Cirrus |
| Mar 19 | 3:02– 3:26 | +67.5 | 1.589 | 0.963 | 36.5 | 108.4 | 276.9 | Snapshot | R,CN | Photometric |
| Mar 24 | 2:33– 6:18 | +72.5 | 1.637 | 1.136 | 36.9 | 100.0 | 275.7 | Snapshot | R,CN | Photometric |
| Mar 25 | 2:29– 7:51 | +73.6 | 1.647 | 1.172 | 36.8 | 98.4 | 275.5 | Snapshot | R,CN | Photometric |
| Mar 28 | 2:43– 7:41 | +76.6 | 1.676 | 1.275 | 36.4 | 94.2 | 275.0 | Imaging | R,OH,CN,BC,C3,C2,GC | Photometric |
| Mar 29 | 2:26– 7:51 | +77.6 | 1.686 | 1.310 | 36.3 | 92.9 | 274.8 | Imaging | R,OH,CN,BC,C3,C2,GC | Photometric |
| Mar 30 | 2:30– 7:00 | +78.6 | 1.696 | 1.344 | 36.1 | 91.6 | 274.7 | Imaging | R,OH,CN,BC,C3,C2,GC | Photometric |
| Mar 31 | 2:33– 7:06 | +79.6 | 1.706 | 1.379 | 35.8 | 90.3 | 274.6 | Imaging | R,OH,CN,BC,C3,C2,GC | Photometric |
| Apr 1 | 2:29– 6:27 | +80.6 | 1.716 | 1.413 | 35.6 | 89.0 | 274.5 | Imaging | R,OH,CN,BC,C3,C2,GC | Clouds |
| Apr 23 | 2:51– 5:39 | +102.5 | 1.945 | 2.144 | 27.9 | 65.0 | 273.8 | Imaging | R,CN,BC,C3,C2,GC | Some cirrus |
| Apr 24 | 2:45– 5:40 | +103.5 | 1.956 | 2.175 | 27.5 | 64.0 | 273.8 | Imaging | R,CN,BC,C3,C2,GC | Photometric |
| Apr 25 | 2:53– 5:02 | +104.5 | 1.967 | 2.206 | 27.1 | 63.1 | 273.8 | Imaging | R,CN | Clouds |
| Apr 26 | 2:57– 5:29 | +105.5 | 1.977 | 2.237 | 28.7 | 62.1 | 273.9 | Imaging | R,CN,BC,C3,C2,GC | Photometric |
| May 14 | 3:40– 4:30 | +123.5 | 2.173 | 2.758 | 19.4 | 45.6 | 274.6 | Imaging | R,CN,BC | Some cirrus |
| May 15 | 3:12– 4:23 | +124.5 | 2.184 | 2.784 | 19.0 | 44.7 | 274.7 | Imaging | R,CN,BC | Cirrus |

[a] All parameters were taken at the midpoint of each night's observations, and all images were obtained with the Hall 42-in (1.1-m) telescope at Lowell Observatory.

[b] Time from perihelion.

[c] Heliocentric distance.

[d] Geocentric distance.

[e] Solar phase angle.

[f] Solar elongation.

[g] Position angle (P.A.) of the Sun.

Table 3. Apparent Rotation Periods Measured from Pairs of Enhanced CN Images

| Image 1 UT Date | | | ΔT (Day) | Image 2 UT Date | | ΔT (Day) | Time Elapsed (Day) | Intervening Cycles | Rotation Period (hr) |
|---|---|---|---|---|---|---|---|---|---|
| 2009 | Feb | 18.34 | +38.70 | Feb | 25.28 | +45.64 | 6.94 | 4 | 41.63 |
| 2009 | Feb | 19.31 | +39.67 | Feb | 24.45 | +44.81 | 5.14 | 3 | 41.11 |
| 2009 | Feb | 20.34 | +40.70 | Feb | 27.32 | +47.68 | 6.98 | 4 | 41.89 |
| 2009 | Feb | 20.45 | +40.81 | Feb | 27.39 | +47.75 | 6.94 | 4 | 41.65 |
| 2009 | Feb | 21.46 | +41.82 | Feb | 28.45 | +48.81 | 6.99 | 4 | 41.91 |
| 2009 | Feb | 24.37 | +44.73 | Feb | 26.13 | +46.49 | 1.76 | 1 | 42.35 |
| 2009 | Feb | 24.42 | +44.78 | Feb | 26.18 | +46.54 | 1.76 | 1 | 42.20 |
| 2009 | Feb | 25.16 | +45.52 | Mar | 2.43 | +50.79 | 5.27 | 3 | 42.16 |
| 2009 | Feb | 25.21 | +45.57 | Mar | 2.47 | +50.83 | 5.26 | 3 | 42.08 |
| 2009 | Feb | 25.42 | +45.78 | Feb | 27.19 | +47.55 | 1.77 | 1 | 42.53 |
| 2009 | Feb | 25.48 | +45.84 | Feb | 27.24 | +47.60 | 1.76 | 1 | 42.35 |
| 2009 | Feb | 28.36 | +48.72 | Mar | 2.11 | +50.47 | 1.75 | 1 | 41.97 |
| 2009 | Feb | 28.40 | +48.76 | Mar | 2.14 | +50.50 | 1.74 | 1 | 41.82 |
| 2009 | Feb | 28.45 | +48.81 | Mar | 2.22 | +50.58 | 1.76 | 1 | 42.35 |
| 2009 | Mar | 1.34 | +49.70 | Mar | 3.10 | +51.46 | 1.76 | 1 | 42.32 |

Table 4. Measured Axis Position Angles from CN Images

| UT Date | | | Axis Position Angle (°)[a] |
|---|---|---|---|
| 2009 | Jan | 30.44 | 290 |
| 2009 | Jan | 31.48 | 286 |
| 2009 | Feb | 1.52 | 289 |
| 2009 | Feb | 2.42 | 285 |
| Mean First Run | | | 288 ± 3 |
| 2009 | Feb | 24.25 | 300 |
| 2009 | Feb | 24.42 | 298 |
| 2009 | Feb | 25.32 | 298 |
| 2009 | Feb | 25.48 | 300 |
| 2009 | Feb | 27.44 | 299 |
| 2009 | Feb | 28.13 | 297 |
| 2009 | Feb | 28.29 | 297 |
| 2009 | Mar | 2.23 | 299 |
| 2009 | Mar | 2.40 | 300 |
| 2009 | Mar | 3.11 | 295 |
| Mean Second Run | | | 298 ± 4 |
| 2009 | Mar | 19.13 | 288 |
| 2009 | Mar | 25.32 | 288 |
| 2009 | Mar | 28.23 | 282 |
| 2009 | Mar | 29.12 | 289 |
| 2009 | Mar | 30.24 | 287 |
| 2009 | Mar | 31.12 | 293 |
| 2009 | Apr | 1.18 | 285 |
| Mean Third Run | | | 287 ± 6 |

[a] The uncertainty of the individual measurements is estimated to be ± 10°

Table 5. Summary of Model Results

| Nucleus Parameters | |
|---|---|
| Pole Obliquity | 97° |
| Pole Orbital Longitude | 120° |
| Pole RA | 81° |
| Pole Dec | +29° |
| Sidereal Rotation Period | 41.45 ± .05 hr |

| Source Region | East | West | Dust[1] |
|---|---|---|---|
| Latitude | −80° | +77° | +10° |
| Longitude | 125° | 245° | 150° |
| Radius | 10° | 20° | 10° |

| Observing Run[2] | CN Terminal Outflow Velocity (km s$^{-1}$) | Distance from nucleus (km) |
|---|---|---|
| 1 | 0.62 | 26,000 |
| 2 | 0.40 | 17,000 |
| 3 | 0.32 | 13,000 |

| Observing Run | Mean Dust Outflow Velocity (km s$^{-1}$) |
|---|---|
| 1 | 0.10 |
| 2 | 0.09 |
| 3 | 0.07 |

[1]An additional source region was necessary to reproduce the dust features.
[2]The timing for each observing run can be seen in Figure 7

Table 6. Photometric Fluxes and Aperture Abundances for Comet C/Lulin (2007 N3)

| UT Date | Aperture Size (arcsec) | Aperture log ρ (km) | log Emission Band Flux (erg cm$^{-2}$ s$^{-1}$) OH | NH | CN | C$_3$ | C$_2$ | log Continuum Flux (erg cm$^{-2}$ s$^{-1}$ Å$^{-1}$) UV | Blue | Green | log M(ρ) (molecule) OH | NH | CN | C$_3$ | C$_2$ |
|---|---|---|---|---|---|---|---|---|---|---|---|---|---|---|---|
| 2008 Jul 31.18 | 97.20 | 4.76 | −10.89 | −12.39 | −10.89 | −10.77 | −11.12 | −13.77 | −13.18 | −13.18 | 32.64 | 29.52 | 30.25 | 29.94 | 29.94 |
| 2008 Jul 31.29 | 126.70 | 4.88 | −10.42 | −12.11 | −10.81 | −10.69 | −10.84 | −13.34 | −13.05 | −13.24 | 33.10 | 29.80 | 30.33 | 30.02 | 30.22 |
| 2009 Feb 26.26 | 126.70 | 4.29 | −8.61 | −10.06 | −9.22 | −9.05 | −9.13 | −11.95 | −11.58 | −11.59 | 32.65 | 30.05 | 30.15 | 29.94 | 30.21 |
| 2009 Feb 26.27 | 204.50 | 4.49 | −8.34 | −9.79 | −8.94 | −8.88 | −8.84 | −11.76 | −11.41 | −11.51 | 32.92 | 30.32 | 30.43 | 30.11 | 30.50 |
| 2009 Feb 26.28 | 126.70 | 4.29 | −8.61 | −10.05 | −9.25 | −9.07 | −9.14 | −11.98 | −11.60 | −11.61 | 32.65 | 30.06 | 30.12 | 29.92 | 30.20 |
| 2009 Feb 26.35 | 204.50 | 4.49 | −8.30 | −9.74 | −8.91 | −8.87 | −8.82 | −11.71 | −11.38 | −11.39 | 32.97 | 30.36 | 30.46 | 30.12 | 30.52 |
| 2009 Feb 26.36 | 62.40 | 3.98 | −9.09 | −10.56 | −9.77 | −9.50 | −9.67 | −12.25 | −11.88 | −11.89 | 32.17 | 29.54 | 29.60 | 29.50 | 29.68 |
| 2009 Feb 26.37 | 126.70 | 4.29 | −8.59 | −10.04 | −9.23 | −9.09 | −9.13 | −11.92 | −11.58 | −11.59 | 32.67 | 30.06 | 30.14 | 29.90 | 30.21 |
| 2009 Feb 26.39 | 204.50 | 4.49 | −8.29 | −9.72 | −8.91 | −8.87 | −8.82 | −11.71 | −11.38 | −11.40 | 32.97 | 30.38 | 30.46 | 30.12 | 30.53 |
| 2009 Feb 26.40 | 155.90 | 4.38 | −8.46 | −9.89 | −9.09 | −9.00 | −8.99 | −11.83 | −11.49 | −11.50 | 32.80 | 30.21 | 30.28 | 30.00 | 30.35 |
| 2009 Feb 26.41 | 97.20 | 4.17 | −8.78 | −10.23 | −9.45 | −9.25 | −9.33 | −12.06 | −11.69 | −11.70 | 32.49 | 29.88 | 29.92 | 29.74 | 30.01 |
| 2009 Feb 26.42 | 62.40 | 3.98 | −9.11 | −10.57 | −9.80 | −9.53 | −9.68 | −12.26 | −11.89 | −11.90 | 32.16 | 29.54 | 29.57 | 29.46 | 29.66 |
| 2009 Mar 21.23 | 97.20 | 4.56 | −9.24 | −11.20 | −9.95 | −9.90 | −9.91 | −13.01 | −12.73 | −12.75 | 32.91 | 29.83 | 30.32 | 30.00 | 30.33 |
| 2009 Mar 21.24 | 97.20 | 4.56 | −9.25 | −11.21 | −9.95 | −9.90 | −9.91 | −13.04 | −12.72 | −12.79 | 32.91 | 29.83 | 30.32 | 29.99 | 30.33 |
| 2009 Mar 21.27 | 97.20 | 4.56 | −9.24 | −11.53 | −9.94 | −9.90 | −9.90 | −13.05 | −12.72 | −12.77 | 32.91 | 29.51 | 30.33 | 30.00 | 30.34 |
| 2009 Mar 21.30 | 48.60 | 4.26 | −9.66 | −11.58 | −10.38 | −10.20 | −10.36 | −13.35 | −13.00 | −13.02 | 32.50 | 29.46 | 29.89 | 29.70 | 29.88 |
| 2009 Mar 21.34 | 97.20 | 4.56 | −9.20 | −11.05 | −9.94 | −9.90 | −9.90 | −13.10 | −12.77 | −12.79 | 32.95 | 29.98 | 30.33 | 29.99 | 30.34 |
| 2009 Apr 21.15 | 77.80 | 4.77 | −10.03 | −12.22 | −10.73 | −10.73 | −10.70 | −13.68 | −13.30 | −13.33 | 32.94 | 29.58 | 30.34 | 29.93 | 30.30 |
| 2009 Apr 21.16 | 126.70 | 4.98 | −9.74 | −12.05 | −10.49 | −10.57 | −10.42 | −13.53 | −13.23 | −13.27 | 33.23 | 29.75 | 30.58 | 30.08 | 30.58 |
| 2009 Apr 21.20 | 77.80 | 4.77 | −9.99 | −12.49 | −10.75 | −10.74 | −10.69 | −13.67 | −13.24 | −13.37 | 32.98 | 29.31 | 30.32 | 29.91 | 30.31 |
| 2009 Apr 21.22 | 126.70 | 4.98 | −9.72 | −12.33 | −10.49 | −10.57 | −10.44 | −13.67 | −13.18 | −13.23 | 33.25 | 29.48 | 30.58 | 30.08 | 30.56 |
| 2009 May 12.18 | 126.70 | 5.09 | −10.13 | — | −10.84 | — | −10.81 | −13.64 | −14.02 | −14.04 | 33.16 | — | 30.57 | — | 30.51 |
| 2009 May 12.19 | 77.80 | 4.88 | −10.39 | — | −11.08 | −11.14 | −11.09 | −14.19 | −13.78 | −13.61 | 32.90 | — | 30.32 | 29.84 | 30.24 |
| 2009 May 13.18 | 62.40 | 4.79 | −10.76 | — | −11.23 | −11.58 | −11.29 | −13.54 | −13.70 | −13.65 | 32.54 | — | 30.19 | 29.42 | 30.04 |

Table 7. Photometric Production Rates for Comet C/Lulin (2007 N3)

| UT Date | ΔT (day) | log r$_H$ (AU) | log ρ (km) | log Q (molecule s$^{-1}$) OH | NH | CN | C$_3$ | C$_2$ | log A(θ)fρ (cm) UV | Blue | Green | log Q H$_2$O |
|---|---|---|---|---|---|---|---|---|---|---|---|---|
| 2008 Jul 31.18 | −163.46 | 0.417 | 4.76 | 28.26 .05 | 25.38 .17 | 25.70 .01 | 25.23 .03 | 25.58 .02 | 2.72 .07 | 2.99 .02 | 3.01 .01 | 28.19 |
| 2008 Jul 31.29 | −163.35 | 0.417 | 4.88 | 28.55 .05 | 25.47 .16 | 25.61 .01 | 25.19 .03 | 25.69 .02 | 3.04 .05 | 3.00 .02 | 2.83 .03 | 28.48 |
| 2009 Feb 26.26 | +46.62 | 0.149 | 4.29 | 28.72 .00 | 26.34 .00 | 26.04 .00 | 25.70 .00 | 26.30 .00 | 3.31 .00 | 3.35 .00 | 3.36 .00 | 28.78 |
| 2009 Feb 26.27 | +46.64 | 0.149 | 4.49 | 28.67 .00 | 26.28 .00 | 26.03 .00 | 25.69 .00 | 26.29 .00 | 3.29 .00 | 3.31 .00 | 3.23 .00 | 28.73 |
| 2009 Feb 26.28 | +46.64 | 0.149 | 4.29 | 28.72 .00 | 26.35 .00 | 26.01 .00 | 25.68 .00 | 26.29 .00 | 3.27 .00 | 3.33 .00 | 3.34 .00 | 28.78 |
| 2009 Feb 26.35 | +46.71 | 0.149 | 4.49 | 28.72 .00 | 26.32 .00 | 26.06 .00 | 25.70 .00 | 26.31 .00 | 3.33 .00 | 3.34 .00 | 3.35 .00 | 28.78 |
| 2009 Feb 26.36 | +46.72 | 0.149 | 3.98 | 28.74 .00 | 26.35 .00 | 25.97 .00 | 25.59 .00 | 26.25 .00 | 3.31 .00 | 3.35 .00 | 3.37 .00 | 28.80 |
| 2009 Feb 26.37 | +46.73 | 0.149 | 4.29 | 28.74 .00 | 26.36 .00 | 26.03 .00 | 25.66 .00 | 26.30 .00 | 3.33 .00 | 3.35 .00 | 3.36 .00 | 28.80 |
| 2009 Feb 26.39 | +46.75 | 0.149 | 4.49 | 28.72 .00 | 26.34 .00 | 26.06 .00 | 25.70 .00 | 26.31 .00 | 3.33 .00 | 3.33 .00 | 3.34 .00 | 28.78 |
| 2009 Feb 26.40 | +46.76 | 0.149 | 4.38 | 28.73 .00 | 26.36 .00 | 26.04 .00 | 25.67 .00 | 26.31 .00 | 3.33 .00 | 3.34 .00 | 3.35 .00 | 28.79 |
| 2009 Feb 26.41 | +46.77 | 0.149 | 4.17 | 28.73 .00 | 26.36 .00 | 25.99 .00 | 25.62 .00 | 26.27 .00 | 3.31 .00 | 3.35 .00 | 3.36 .00 | 28.79 |
| 2009 Feb 26.42 | +46.78 | 0.149 | 3.98 | 28.72 .00 | 26.35 .00 | 25.94 .00 | 25.55 .00 | 26.23 .00 | 3.30 .00 | 3.34 .00 | 3.36 .00 | 28.78 |
| 2009 Mar 21.23 | +69.59 | 0.207 | 4.56 | 28.61 .00 | 25.75 .02 | 25.87 .00 | 25.50 .01 | 26.07 .00 | 2.87 .00 | 2.82 .01 | 2.82 .01 | 28.64 |
| 2009 Mar 21.24 | +69.60 | 0.207 | 4.56 | 28.61 .00 | 25.74 .02 | 25.87 .00 | 25.49 .00 | 26.07 .00 | 2.83 .00 | 2.83 .01 | 2.78 .01 | 28.64 |
| 2009 Mar 21.27 | +69.63 | 0.207 | 4.56 | 28.62 .00 | 25.42 .03 | 25.88 .00 | 25.50 .01 | 26.08 .00 | 2.82 .01 | 2.82 .01 | 2.80 .01 | 28.65 |
| 2009 Mar 21.30 | +69.66 | 0.207 | 4.26 | 28.67 .01 | 25.87 .00 | 25.88 .00 | 25.48 .01 | 26.07 .00 | 2.83 .02 | 2.85 .00 | 2.85 .01 | 28.70 |
| 2009 Mar 21.34 | +69.70 | 0.207 | 4.56 | 28.65 .01 | 25.90 .03 | 25.87 .00 | 25.49 .01 | 26.07 .01 | 2.77 .04 | 2.78 .02 | 2.78 .01 | 28.68 |
| 2009 Apr 21.15 | +100.51 | 0.284 | 4.77 | 28.41 .01 | 25.26 .09 | 25.66 .01 | 25.23 .02 | 25.81 .01 | 2.75 .04 | 2.80 .02 | 2.79 .02 | 28.40 |
| 2009 Apr 21.16 | +100.52 | 0.284 | 4.98 | 28.40 .00 | 25.11 .10 | 25.63 .01 | 25.25 .02 | 25.82 .01 | 2.69 .05 | 2.66 .03 | 2.64 .02 | 28.39 |
| 2009 Apr 21.20 | +100.56 | 0.284 | 4.77 | 28.45 .01 | 24.99 .19 | 25.64 .01 | 25.22 .02 | 25.83 .01 | 2.76 .05 | 2.86 .02 | 2.75 .02 | 28.44 |
| 2009 Apr 21.22 | +100.58 | 0.284 | 4.98 | 28.42 .02 | 24.84 .26 | 25.62 .01 | 25.25 .03 | 25.80 .01 | 2.55 .11 | 2.71 .03 | 2.68 .03 | 28.41 |
| 2009 May 12.18 | +121.54 | 0.333 | 5.09 | 28.21 .11 | — | 25.50 .02 | — | 25.63 .02 | 2.78 .18 | 2.08 .23 | 2.08 .16 | 28.18 |
| 2009 May 12.19 | +121.55 | 0.333 | 4.88 | 28.25 .28 | — | 25.52 .05 | 25.04 .15 | 25.63 .03 | 2.45 .40 | 2.53 .13 | 2.72 .06 | 28.22 |
| 2009 May 13.18 | +122.54 | 0.335 | 4.79 | 28.03 .23 | — | 25.52 .02 | 24.68 .16 | 25.57 .04 | 3.20 .08 | 2.71 .08 | 2.79 .05 | 28.00 |

Table 8. Continuum Fluxes and $A(\theta)f\rho$ values from Narrowband Imaging for Comet C/Lulin (2007 N3)

| UT Date | log $r$ | Phase Angle (°) | Phase Adj. log $A(0°)f\rho^a$ | log Continuum Flux (erg cm$^{-2}$ s$^{-1}$ Å$^{-1}$)[b,c] | | log $A(\theta)f\rho$ (cm)[b,c] | |
|---|---|---|---|---|---|---|---|
| | | | | Blue | Red | Blue | Red |
| 2009 Jan 30 | 0.097 | 50.5 | +0.47 | –12.25 | –12.27 | 3.00 | 3.16 |
| 2009 Jan 31 | 0.099 | 50.7 | +0.47 | –12.22 | –12.25 | 3.00 | 3.17 |
| 2009 Feb 1 | 0.100 | 50.9 | +0.47 | –12.21 | –12.23 | 2.96 | 3.13 |
| 2009 Feb 2 | 0.101 | 50.9 | +0.47 | –12.22 | –12.23 | 2.98 | 3.16 |
| 2009 Feb 24 | 0.145 | 8.4 | +0.14 | –11.34 | –11.40 | 3.24 | 3.37 |
| 2009 Feb 28 | 0.154 | 7.9 | +0.13 | –11.49 | – | 3.17 | – |
| 2009 Mar 28 | 0.224 | 36.4 | +0.42 | –12.81 | – | 2.91 | – |
| 2009 Mar 29 | 0.227 | 36.3 | +0.42 | –12.99 | – | 2.76 | – |
| 2009 Mar 30 | 0.229 | 36.1 | +0.42 | –13.04 | – | 2.74 | – |
| 2009 Mar 31 | 0.232 | 35.8 | +0.42 | –13.01 | – | 2.80 | – |
| 2009 Apr 23 | 0.289 | 27.9 | +0.37 | –13.56 | – | 2.75 | – |
| 2009 Apr 24 | 0.291 | 27.5 | +0.36 | –13.56 | – | 2.77 | – |
| 2009 Apr 26 | 0.296 | 28.7 | +0.37 | –13.53 | – | 2.82 | – |

[a] Adjustment to 0° solar phase angle to $A(\theta)f\rho$ values based on assumed phase function (see text)
[b] Flux and $A(\theta)f\rho$ values are for apertures of log $\rho$ = 4.60 km
[c] The uncertainties are estimated to be less than 5% and are dominated by the absolute calibrations

Figure 1

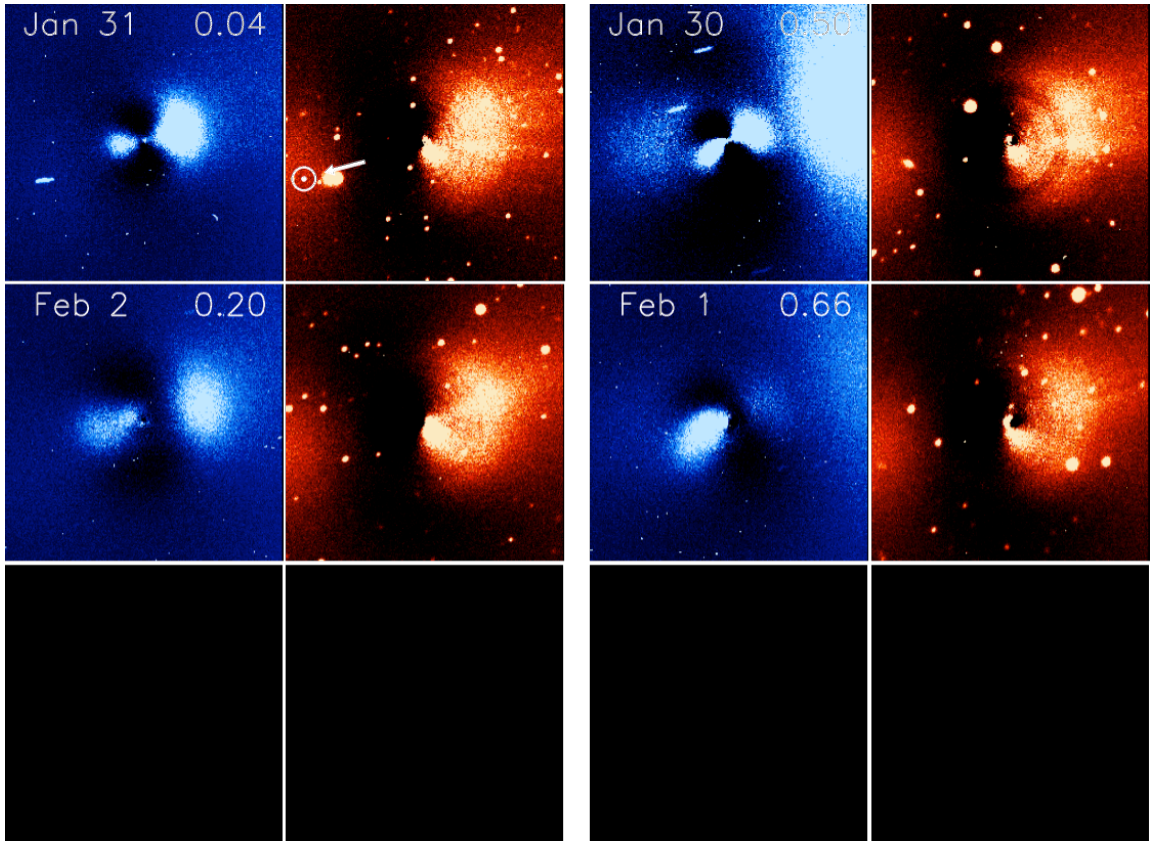

Figure 1. A rotational sequence of enhanced CN and *R*-band (dust) images from our first imaging run. Time steps are ~0.15 in phase, or around 6 hours apart in the rotational cycle. Rotational phase is indicated in the upper right corner of each image; blank panels indicate a rotational phase for which we have no observations. All images are 200,000 km across at the comet, with north up and east to the left. The direction to the Sun is indicated in the first panel and barely changes during this time frame.

Figure 2

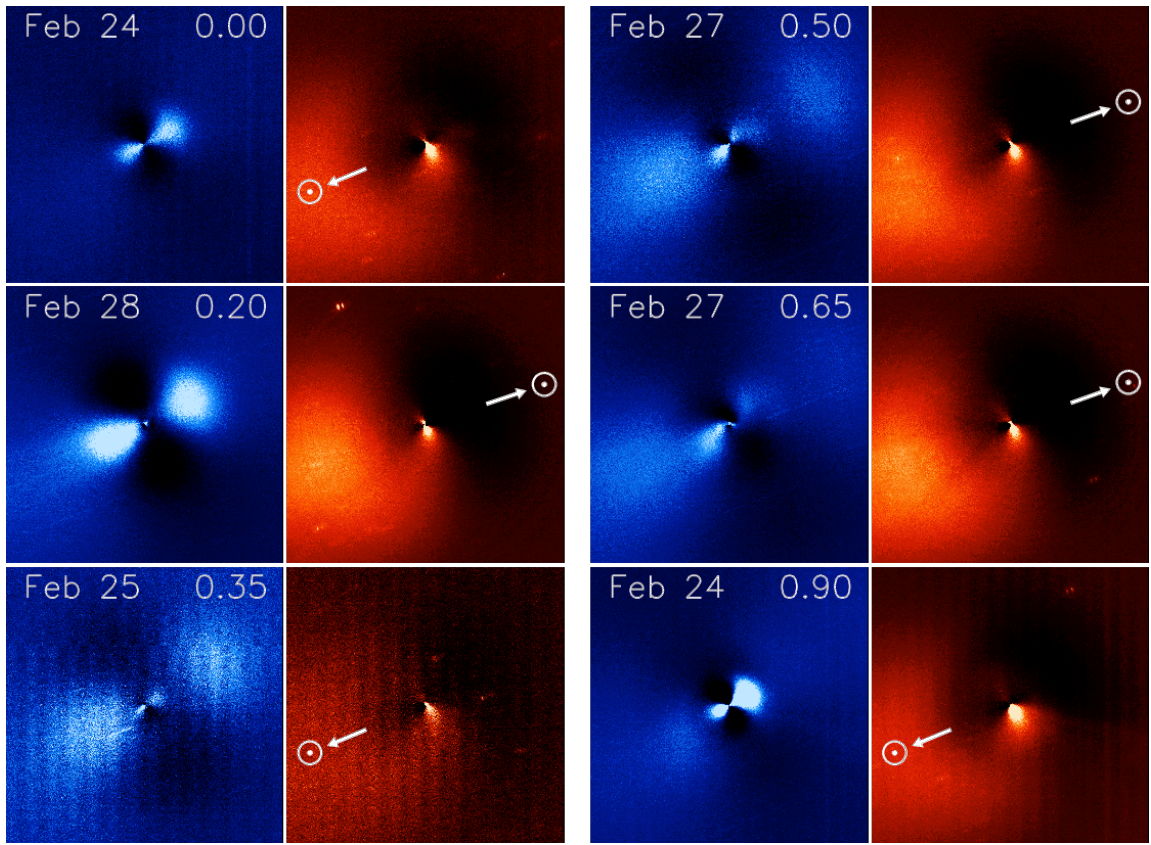

Figure 2. A rotational sequence of enhanced CN and *R*-band (dust) images from our second imaging run. The direction to the Sun is indicated in each panel. All other details are as given in Fig 1.

Figure 3

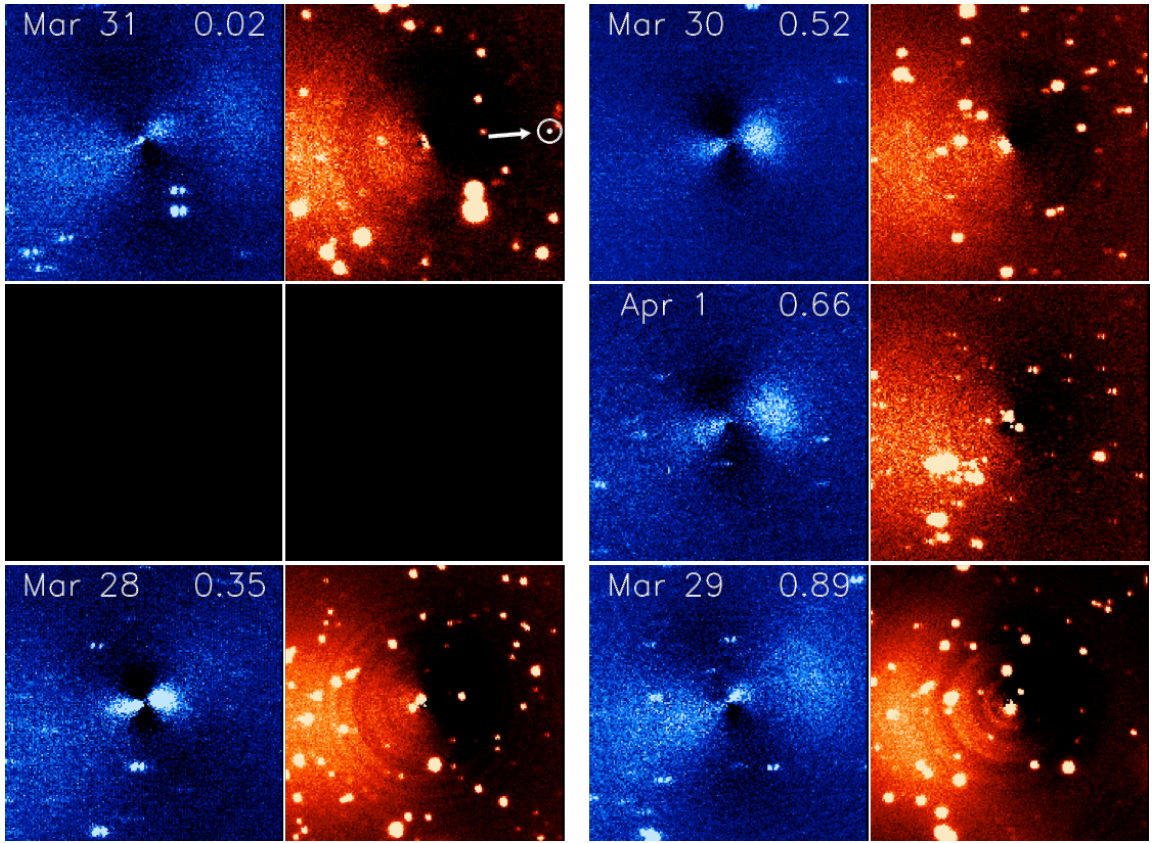

Figure 3. A rotational sequence of enhanced CN and *R*-band (dust) images from our third imaging run. The direction to the Sun is indicated in the first panel and barely changes during this time frame. All other details are as given in Fig 1.

Figure 4

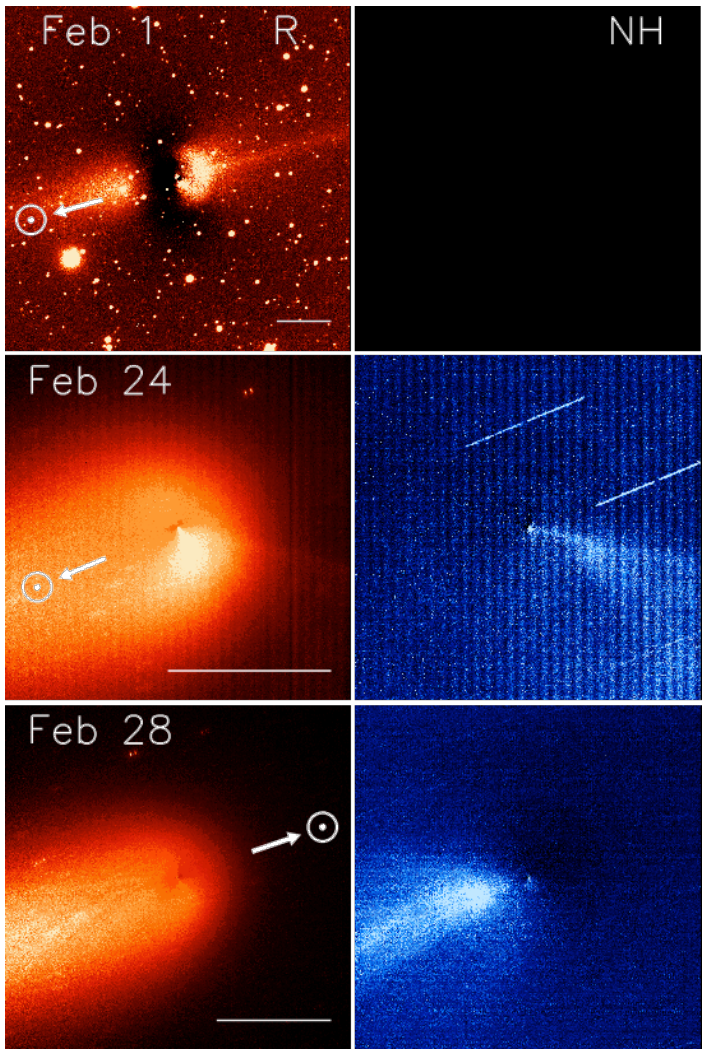

Figure 4. The ion features observed in enhanced images from the *R*-band (dust; left column) and NH filters (right column). The ion feature in the R-band images, which we believe to be $H_2O^+$, appears at a PA of 285° on Feb 1 (top left), 260° on Feb 24 (middle left), and 120° on Feb 28 (bottom left). The ion feature in the NH filters on the last two nights (middle right and bottom right), having the same PAs, is believed to be $OH^+$. These ion features abruptly change directions between Feb 24 and Feb 28, an interval exactly surrounding minimum phase angle. A scale bar 100,000 km long and the direction to the Sun are indicated for each date on the *R*-band images.

Figure 5 (Top)

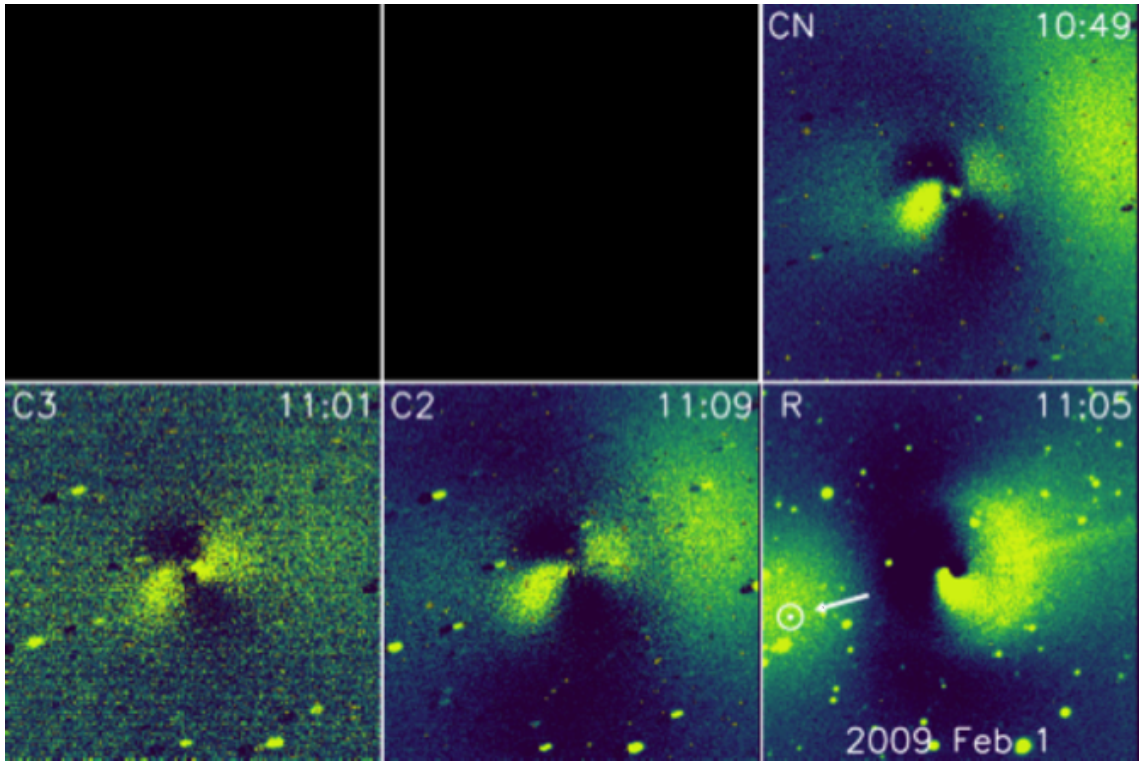

Figure 5. Enhanced narrowband gas and *R*-band (dust) features observed in Comet Lulin during our three main observing runs. Blank panels are shown when we were not able to observe with a specific filter due to low signal-to-noise. All images are 250,000 km across at the comet. A night from our first observing run is shown in the top panel, from our second run in the middle panel, and from our third run in the bottom panel. The structure of OH is very different from that of the carbon-bearing species, evidence that a different driver may exist such as CO or $CO_2$ for the near-polar jets.

Figure 5 (Middle)

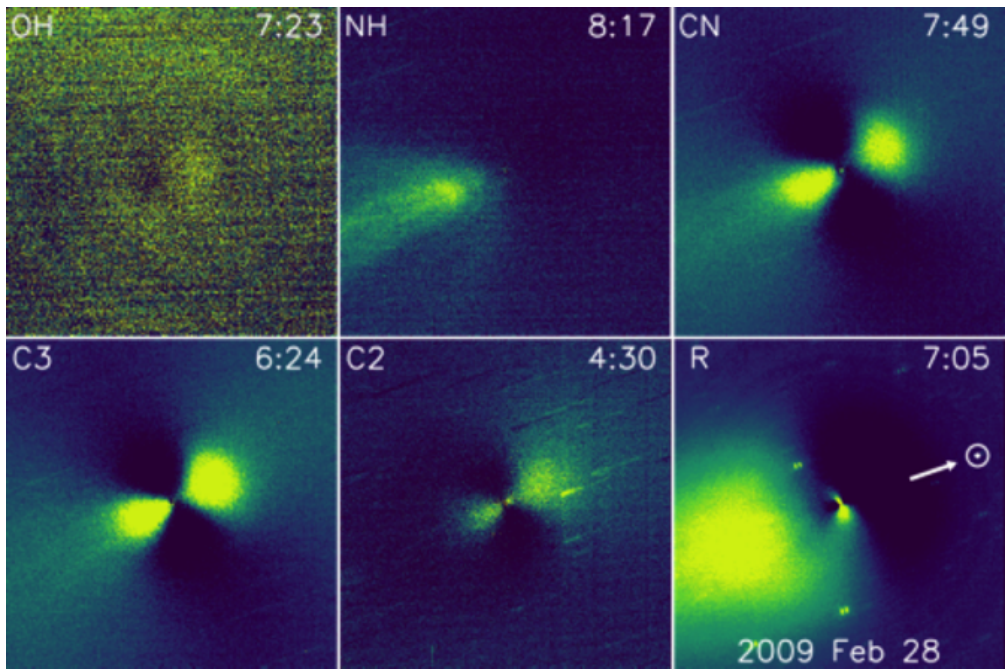

Figure 5 (Bottom)

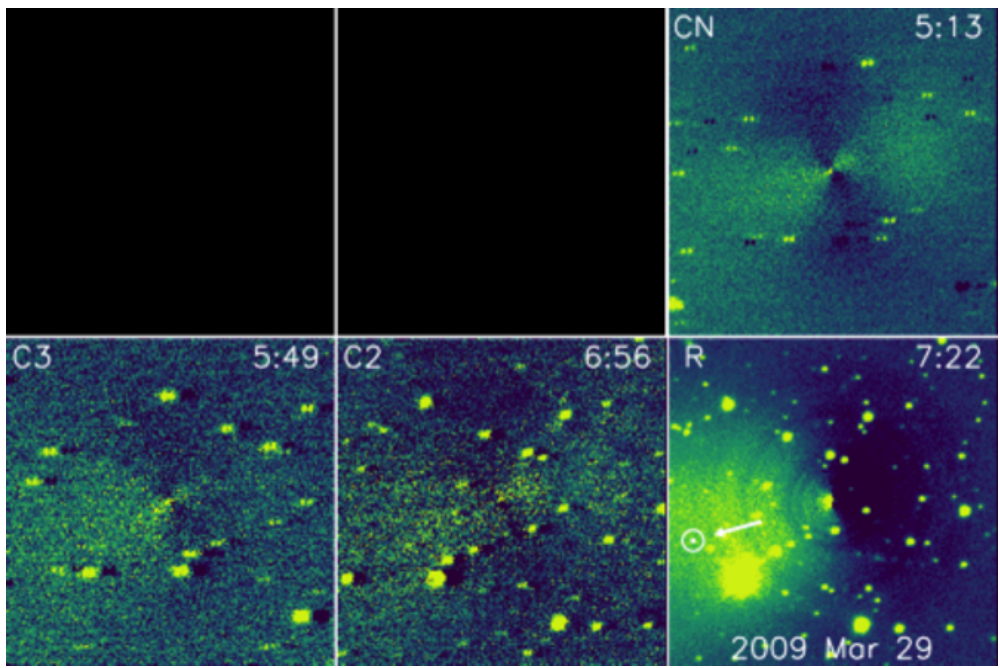

Figure 6

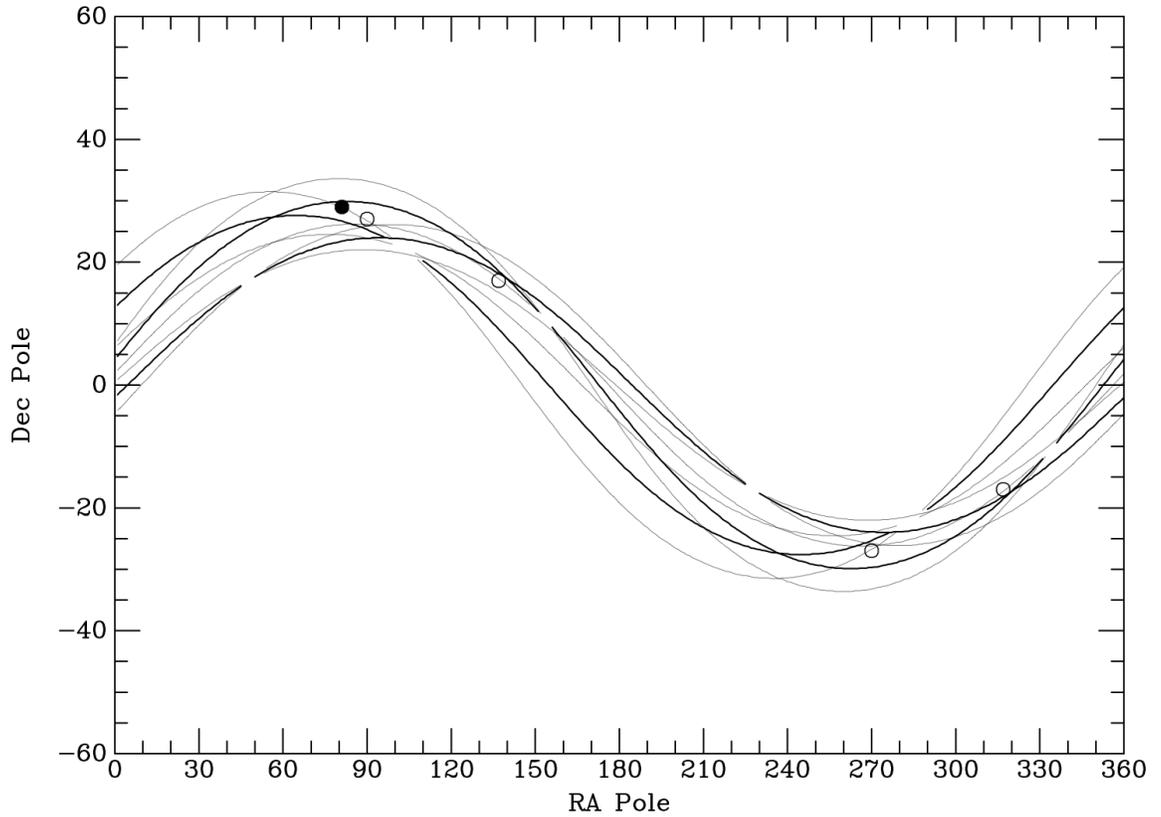

Figure 6. The potential rotational pole solutions, generated from our mean axis measurements listed in Table 4. Each thick line is the great circle defining possible pole solutions, and the thinner lines are the estimated uncertainty in the solutions. The solution for the tilt of the pole is fairly precise, while the position for the principle angle has a larger margin for potential solutions. The open circles mark our preliminary potential solutions, while the filled circle indicates our final answer, derived from matching the jet shapes and locations between the images and the model for each observing run.

Figure 7

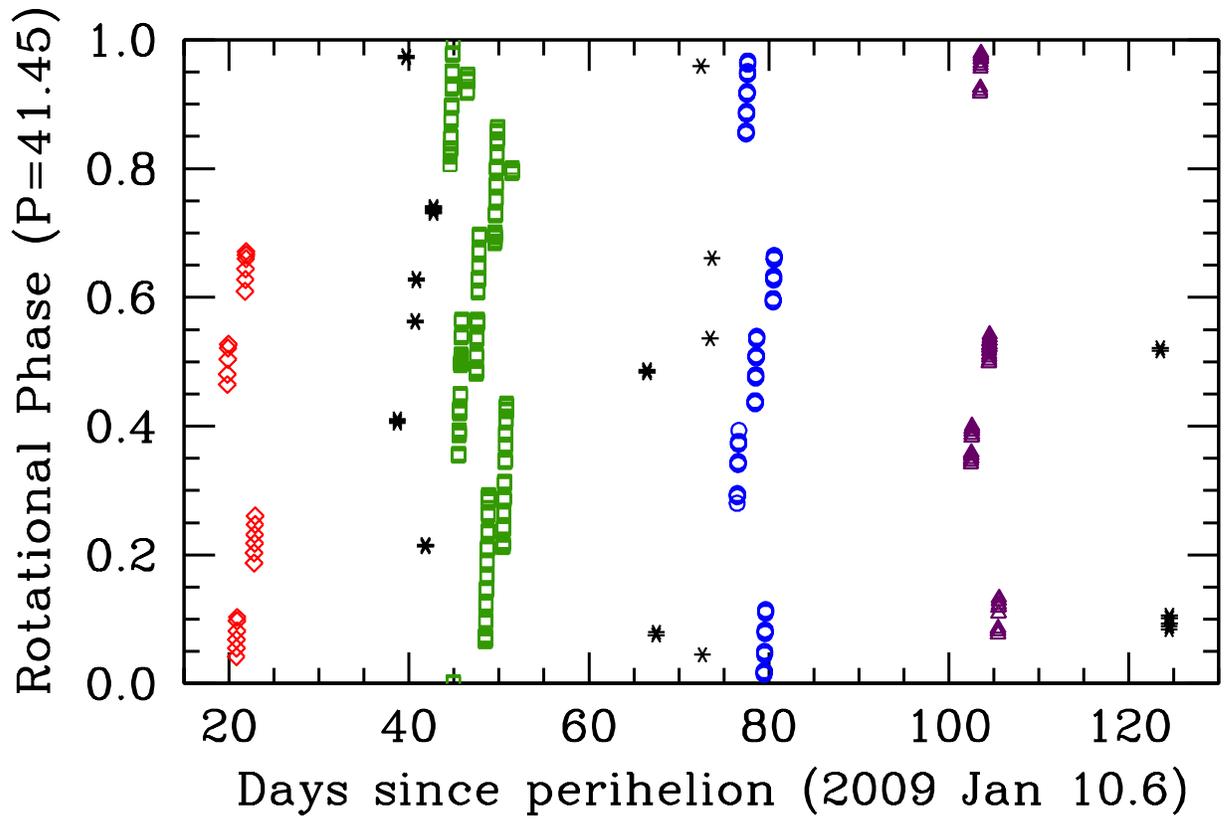

Figure 7. The rotational phase coverage of our CN images as a function of days since perihelion ($\Delta T$), using the sidereal rotation period of 41.45 hr (discussed in Section 4.5). Our first observing run is represented by red diamonds, the second by green squares, the third by blue circles, and the fourth by purple triangles. Snapshot images from between the main imaging runs are shown as black stars.

Figure 8

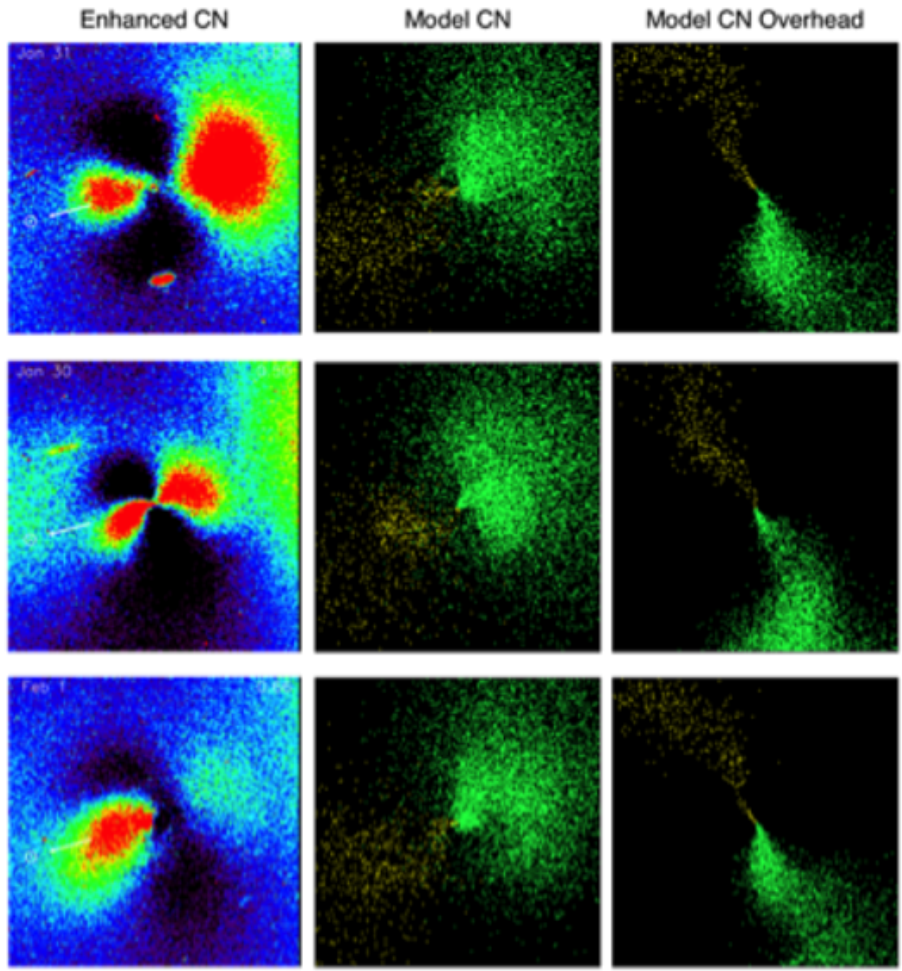

Figure 8: The most diagnostic enhanced CN images from our first observing run (left column), compared with the concurrent times from our model solution as seen from Earth (middle column), and from above the comet, i.e. perpendicular to our line of sight (right column). The position and the motion of the jets work well, though the relative brightness of the eastern jet to the western jet in the model is lower than what was observed. Any adjustments to the model that make the east jet appear stronger here compromise our overall solution. There are strong projection effects during this time frame, as can be seen in the overhead view of the comet in the right panels. The sub-Earth latitude during this time frame is steady at +59°±1°. All panels are scaled to 120,000 km across at the comet.

Figure 9

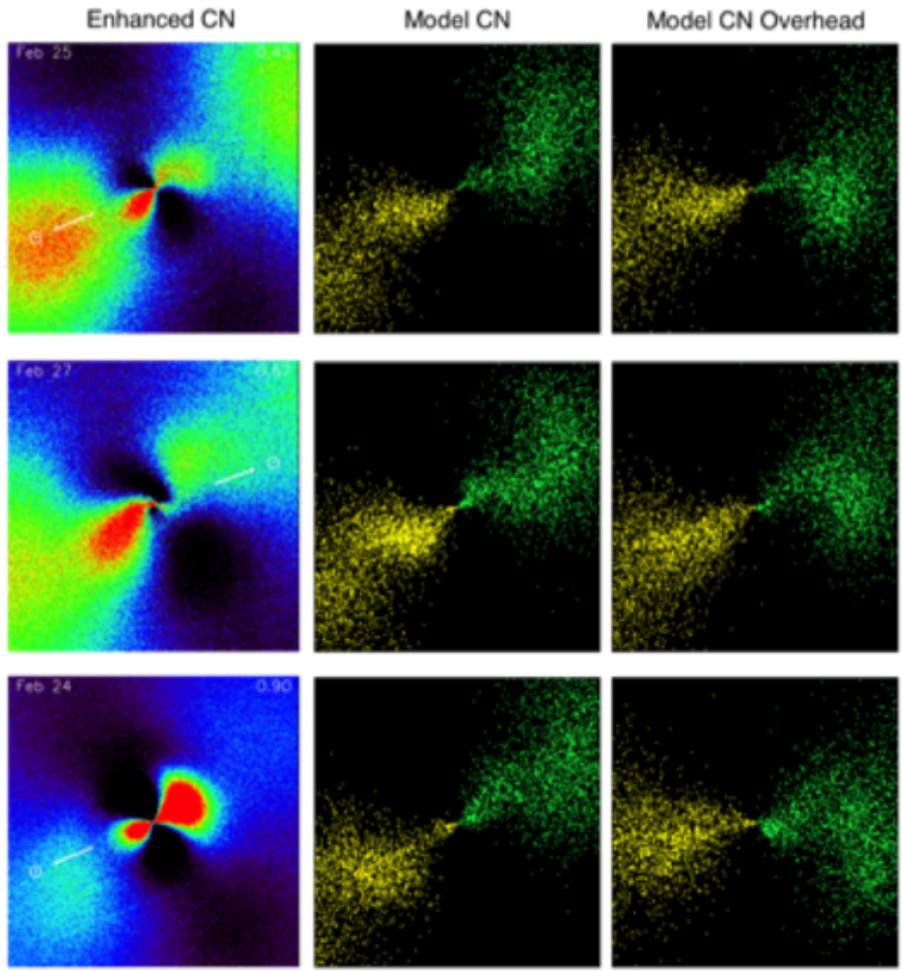

Figure 9: The most diagnostic enhanced CN images and model solutions from our second observing run (left column), compared with the concurrent times from our model solution as seen from Earth (middle column), and from above the comet, i.e. perpendicular to our line of sight (right column). We are seeing the comet side-on during this time frame, with minimal projection effects. The sub-Earth latitude during this time frame changes from -3° on Feb 24 to -20° on Feb 27. All panels are scaled to 120,000 km across at the comet.

Figure 10

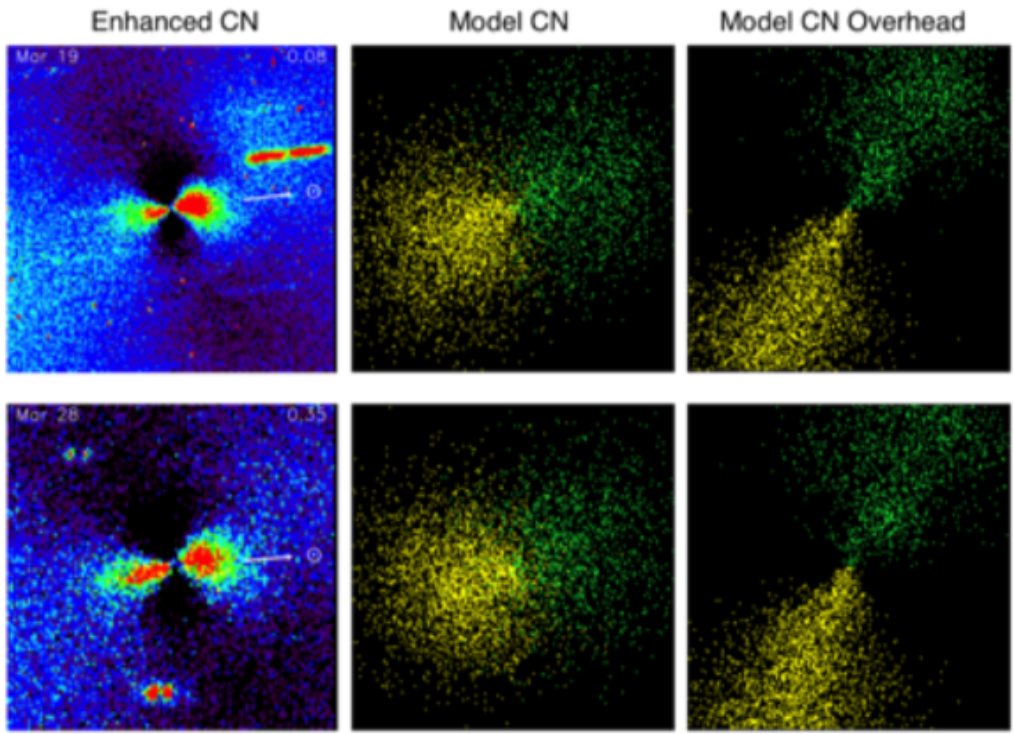

Figure 10: The most diagnostic enhanced *CN* images and model solutions from before (Mar 19) and during (Mar 28) our third observing (left column), compared with the concurrent times from our model solution as seen from Earth (middle column), and from above the comet, i.e. perpendicular to our line of sight (right column). The first image was taken significantly earlier than the second, when the jets were more distinguishable and before our view of the jets became nearly pole-on, as can be seen in the top-town view. The sub-Earth latitude was at -64° on Mar 19, and decreased to -69° by Mar 28. All panels are scaled to 120,000 km across at the comet.

Figure 11

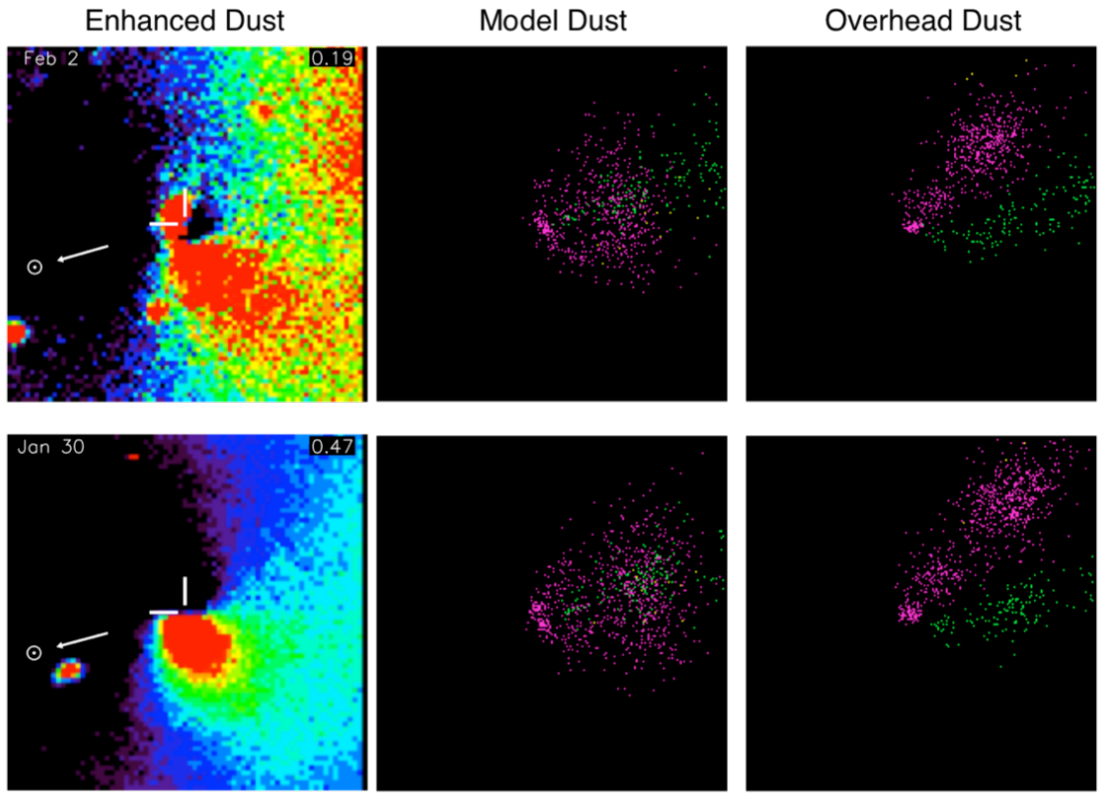

Figure 11: The most diagnostic enhanced *R*-band (dust) images from our first observing run (left column), compared with the concurrent times from our model solution as seen from Earth (middle column), and from above the comet, i.e. perpendicular to our line of sight (right column). Our model focuses on the innermost dust features, where we can see rotational change in both the images and our model representation. All panels are scaled to 60,000 km across at the comet.

Figure 12

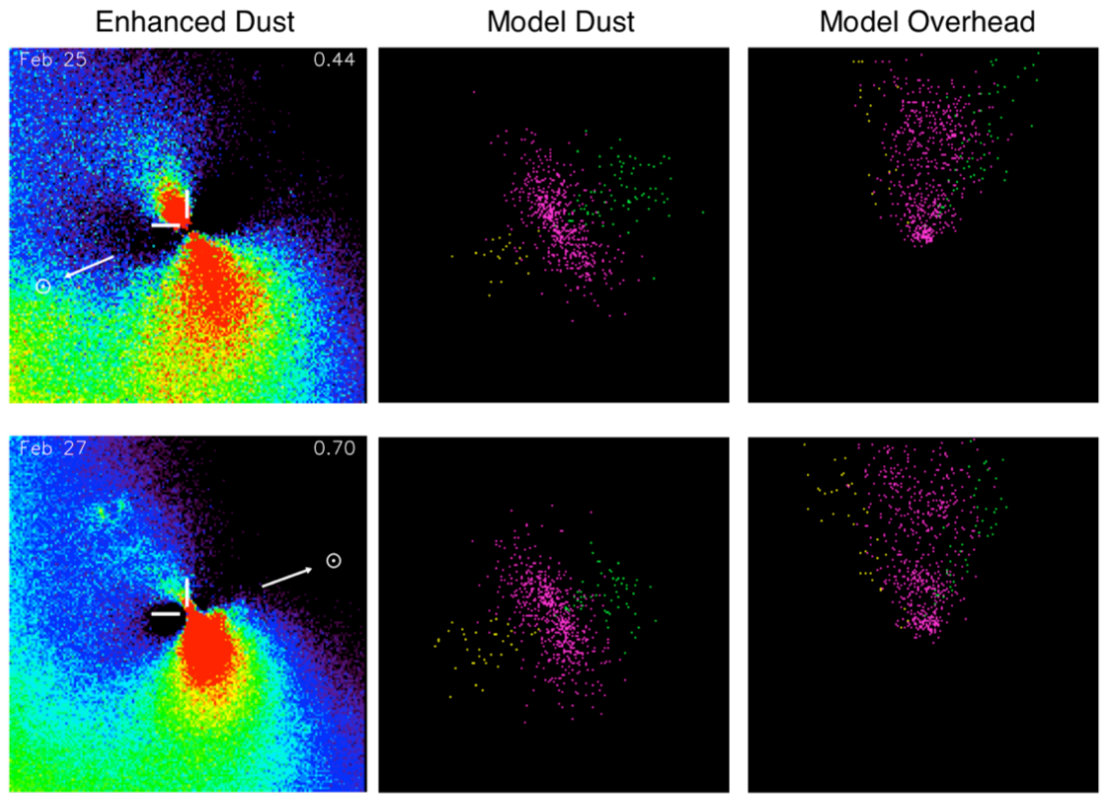

Figure 12: The most diagnostic enhanced *R*-band (dust) images from our second observing run (left column), compared with the concurrent times from our model solution as seen from Earth (middle column), and from above the comet, i.e. perpendicular to our line of sight (right column). Our model focuses on the innermost dust features, where we can see the relative brightness change between the top and bottom features. Our view is in the plane of this near-equatorial source area, and we are seeing the rotational variability close to the nucleus. All panels are scaled to 60,000 km across at the comet.

Figure 13

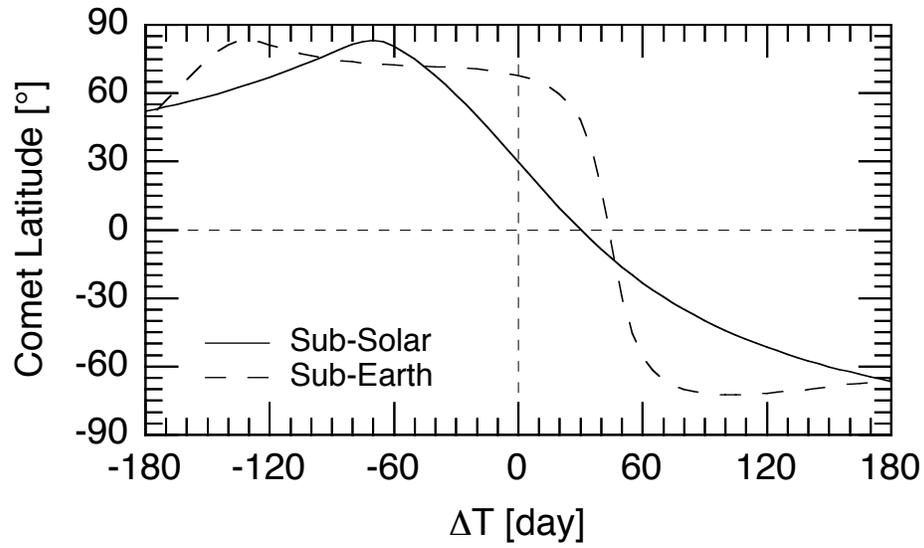

Figure 13: The sub-solar and sub-Earth latitudes as a function of time ($\Delta T$), given our derived pole solution from the CN model. Note the relatively rapid change during our observing intervals, spanning $\Delta T = +19$ to $\Delta T = +81$ day.

Figure 14

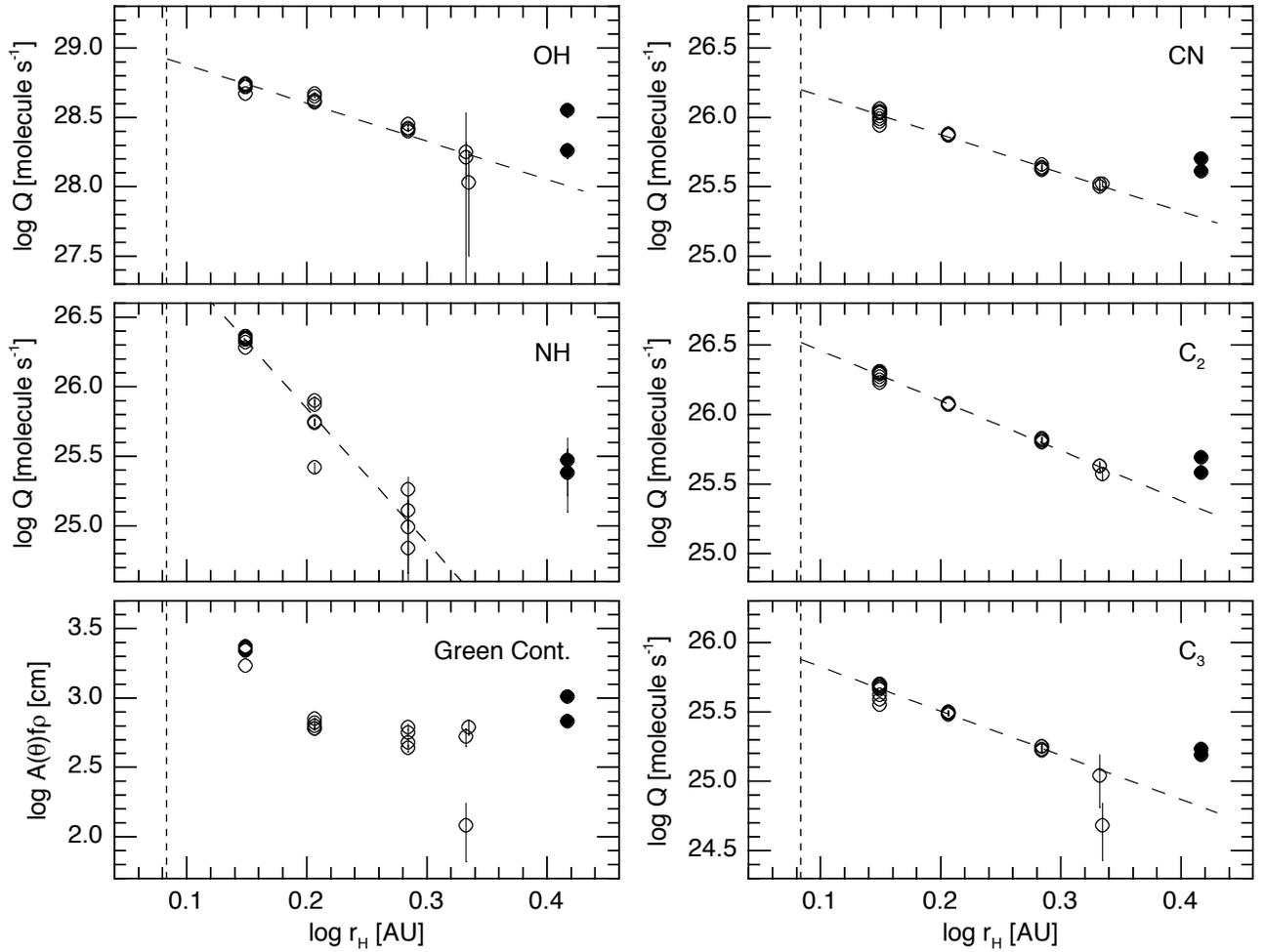

Figure 14: The logarithmic production rates for Comet Lulin as a function of the log of the heliocentric distance. Filled symbols represent data obtained before perihelion, while open symbols represent the post-perihelion data. 1-σ uncertainties are also plotted. The vertical dashed line represents the perihelion distance.

Figure 15

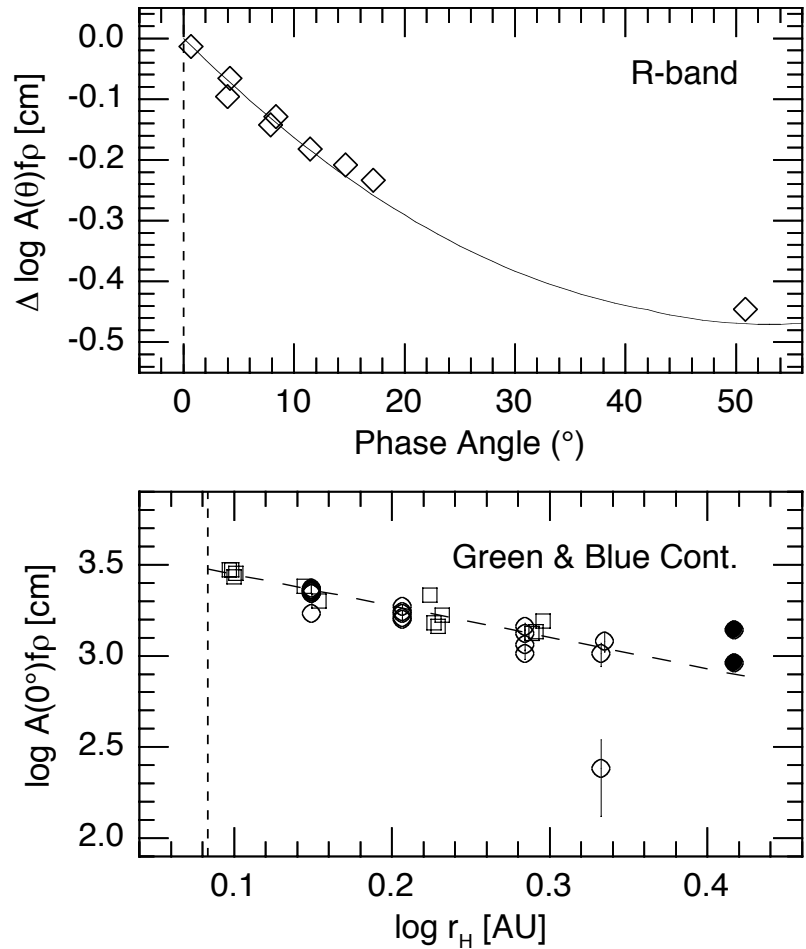

Figure 15: The extracted log $A(\theta)f\rho$ values from imaging vs phase angle (top) and the phase adjusted log $A(0°)f\rho$ photometry production rates vs the log of the heliocentric distance (bottom). For the top panel, each data point was calculated from the fluxes extracted from *R*-band images during our second observing run, when the phase angle decreased from 9° down to 0.7° and back up to 17°, along with one point from our first run. We removed the nominal $r_H$-dependence of -1.74 that we previously found, and see that Lulin's phase curve is an extremely good match to our standard composite phase curve (Schleicher & Bair 2011). In the bottom panel, the phase-adjusted $Af\rho$ results for the green continuum (circles) are overlaid with the blue continuum values from the narrowband imaging (squares). Excluding the pre-perihelion points (filled circles), note the flat and shallow slope as compared to the unadjusted values in Figure 14. The slope of the dust, -1.74±0.1, is much more shallow than any of the gas species.